\documentclass[final, 3p]{elsarticle}

\usepackage{amssymb}
\usepackage{multirow}
\usepackage{siunitx}
\usepackage{url}
\usepackage{amsmath}
\usepackage{graphicx}
\usepackage[table,xcdraw]{xcolor}
\usepackage{subcaption}
\usepackage[export]{adjustbox}
\usepackage[title]{appendix}
\usepackage{array,booktabs,arydshln,xcolor}
\usepackage{colortbl}
\usepackage{caption}
\usepackage{longtable}

\defcitealias{UNreport2022}{UN, 2022}
\defcitealias{EPA2021}{EPA, 2021}
\defcitealias{McKinsey2021}{2021}
\defcitealias{cambridgemaGIS}{City of Cambridge, 2023}

\usepackage{floatrow}
\newfloatcommand{capbtabbox}{table}[][\FBwidth]

\usepackage{blindtext}

\usepackage{natbib}

\usepackage[markup=underlined]{changes}

\usepackage{enumitem}
\setlist[enumerate]{label=\textit{\arabic*.}}

\begin{document}

\begin{frontmatter}

\title{Shared lightweight autonomous vehicles for urban food deliveries: A simulation study}

\author[inst1]{Ainhoa Genua Cerviño}
\author[inst1]{Naroa Coretti Sanchez \footnote{Corresponding author naroa@mit.edu}}
\author[inst1]{Elaine Liu Wang}
\author[inst1]{Arnaud Grignard}
\author[inst1]{Kent Larson}

\affiliation[inst1]{MIT Media Lab, Cambridge, USA}

\begin{abstract}
In recent years, the rapid growth of on-demand deliveries, especially in food deliveries, has spurred the exploration of innovative mobility solutions. In this context, lightweight autonomous vehicles have emerged as a potential alternative. However, their fleet-level behavior remains largely unexplored. To address this gap, we have developed an agent-based model and an environmental impact study assessing the fleet performance of lightweight autonomous food delivery vehicles. This model explores critical factors such as fleet sizing, service level, operational strategies, and environmental impacts. We have applied this model to a case study in Cambridge, MA, USA, where results indicate that there could be environmental benefits in replacing traditional car-based deliveries with shared lightweight autonomous vehicle fleets. Lastly, we introduce an interactive platform that offers a user-friendly means of comprehending the model's performance and potential trade-offs, which can help inform decision-makers in the evolving landscape of food delivery innovation.

\end{abstract}


\end{frontmatter}

\section{Introduction}

Over the past century, the world has undergone substantial transformations. Global population has increased year-by-year, and this trajectory is projected to continue, reaching the 9.7 billion mark by 2050 \citep{UNreport2022}. This would entail a growth of 1.7 billion people relative to 2022, with most of this growth being concentrated in urban areas \citep{UNreport2022}. This demographic shift will pose diverse challenges in cities, including a surge in urban mobility demand and subsequent traffic-related problems, which can lead to adverse environmental and socioeconomic outcomes \citep{li2019challenges}.

In the United States (US), a significant share of the rising urban mobility demand is attributed to on-demand deliveries. While in 2009, online sales represented only 4\% of total US retail sales, by 2019, this figure had risen to 11\%, and by 2021, it had reached 15\%  \citep{jaller2023commerce}. E-commerce, while potentially reducing the number of trips to physical stores, tends to encourage more frequent purchases, resulting in complex and extended delivery routes \citep{mokhtarian2004conceptual, pettersson2018commerce}. Consequently, the environmental implications of e-commerce have been found to have a high dependency on factors such as demand consolidation, delivery network, vehicle types, and return rates \citep{mangiaracina2015review}.


Within the sector of e-commerce deliveries, food delivery services have exhibited remarkably rapid growth. According to McKinsey \& Co  \citep{McKinsey2021}, the market grew four to seven fold between 2018 and 2021, with a global market estimated to be worth more than \$150 billion. This exponential market growth has led to a surge in the exploration of innovative mobility solutions for food deliveries. 

Academic and industrial players are currently exploring lightweight autonomous vehicles that would provide an on-demand food delivery service. For instance, companies like DoorDash have established DoorDash Labs to develop automation and robotics systems for last-mile logistics\footnote{https://doordash.news/company/introducing-doordash-labs-doordashs-robotics-and-automation-arm/}. Similarly, Uber Eats has recently partnered with Nuro, a startup working on autonomous delivery vehicles\footnote{https://www.washingtontimes.com/news/2022/sep/13/uber-eats-partners-with-nuro-for-driverless-delive/}, and Amazon has developed an autonomous delivery system, the Amazon Scout\footnote{https://www.aboutamazon.com/news/transportation/whats-next-for-amazon-scout}. Concurrently, academia is exploring an alternative approach, focusing on multi-functional shared lightweight autonomous vehicles capable of serving as a mobility-on-demand system for people during peak hours and transition into package or food delivery during periods of reduced user demand \citep{sanchez2020autonomous,lin2021affordable}.

These innovative vehicles hold the promise of delivering several advantages. Firstly, they offer the potential to facilitate a transition to lighter vehicles that are more suitable for delivering small food packages due to their lower environmental emissions. Secondly, these shared lightweight autonomous vehicles (SLAV) are purposefully designed to operate on sidewalks or bike lanes, eliminating the necessity for additional road infrastructure in urban areas, which could, in turn, support the shift toward more people-centric and less car-centric urban environments \citep{poeting2019simulation}.

However, despite the substantial efforts invested in industry and academia to develop lightweight autonomous systems for food deliveries, there is a research gap concerning their performance at the fleet level. Notably, a comprehensive review of agent-based models on autonomous robots up to 2020 by \citet{li2021systematic} highlights the need for more logistics-related studies, underscoring the need for further research in this field. The significance of this research gap is particularly evident when considering prior studies that highlight the crucial role of fleet-level investigations in understanding the performance and environmental effects of emerging mobility systems \citep{fagnant2014travel,chen2016operations,sanchez2022performance,romano2021simulation}. 

To bridge this gap, this paper presents three fundamental contributions: 1) an agent-based model that simulates the behavior of lightweight autonomous fleets for food deliveries, 2) an environmental impact assessment of the shift from car-based deliveries to fleets of shared lightweight autonomous vehicles, and 3) an interactive simulation tool that offers user-friendly means of understanding the model's performance and potential trade-offs. In a process that incorporates realistic data on food deliveries and considers different design parameters and operational strategies, our study provides an extensive analysis evaluating the performance and potential implications of these new systems. In particular, the outcomes of this study provide insights into essential parameters related to food delivery fleets, encompassing aspects such as optimal fleet size, service level, operational strategies, and environmental impacts.

The remainder of the paper is structured as follows. The first two subsections (Section \ref{subsec:Literature review} and Section \ref{subsec:Contribution}) address the literature review and contribution of this paper. Section \ref{sec:Modeling} presents the details of the modeling approach. Section \ref{sec:Experimental setup} shows the experimental setup of the defined model. Section \ref{sec:Results and discussion} gathers and discusses the obtained results. Next, Section \ref{sec:Tangible interface} presents a tangible simulation tool that allows the exploration of the model's outcomes in an interactive way.  Finally, Section \ref{sec:Conclusions} summarizes study's main conclusions.

\subsection{Literature review}
\label{subsec:Literature review}

The field of shared autonomous micro-mobility encompasses the use of shared lightweight autonomous vehicles as a mobility on-demand system. Numerous studies have examined the fleet-level behavior within this field, focusing on aspects such as service quality and environmental impacts \citep{sanchez2022simulation,sanchez2022performance,sanchez2022can,kondor2021estimating}. However, the literature is relatively scarce when it comes to utilizing shared lightweight autonomous vehicles for logistics.  \citet{li2021systematic} review all the Agent-Based Models (ABMs) published in this field until 2020 and highlight the limited number of relevant studies in the field of autonomous robots for deliveries, emphasizing the need for further research. Out of a total of four studies that they identify, two focus on package deliveries \citep{poeting2019simulation,haas2017developing} and only the other two are related to food deliveries \citep{samouh2020multimodal,de2019autonomous}. On one hand, \citet{samouh2020multimodal} present an ABM for food delivery considering three scenarios involving drones, ground robots, and a combination of both. They analyze the fleet sizes needed to answer a specific ad-hoc developed demand profile during a particular hour of the day. On the other hand, \citet{de2019autonomous}, instead, study the use of autonomous sidewalk robots to serve food delivery trips. 

The research presented in this article differs from such studies in several aspects. First, regarding the input demand dataset, our model is based on a fine-grained food delivery demand dataset with high-resolution time and location information (see Section \ref{sec:demand}). The aforementioned studies, instead, utilize simplistic datasets such as the one-hour uniform distribution of orders in the case of \citet{samouh2020multimodal} or the random and uniform distribution of customers and restaurants in the case of \citet{de2019autonomous}. Realistic demand patterns produce more accurate results and, therefore, our research poses a significant step forward in this sense \citep{li2021systematic}. Secondly, our model allows for the analysis of the performance of lightweight autonomous systems across various vehicle configurations and charging operational strategies and enables a comparison to a baseline scenario represented by cars. Previous studies, instead, study simpler vehicle configurations and lack a baseline that allows for a comparison \citep{samouh2020multimodal,de2019autonomous}. Lastly, unlike previous studies, our work includes an environmental impacts analysis, which is an increasingly important aspect in evaluating the performance of mobility systems \citep{tiboni2021urban}.

In light of previous work, it can be concluded that there is still a relevant literature gap in developing an in-depth study of the fleet-level performance and environmental impacts of shared lightweight autonomous vehicles for food deliveries. 

\subsection{Contribution}
\label{subsec:Contribution}

In order to fill the existing literature gap, this study aims to comprehensively examine the use of lightweight autonomous systems for food deliveries. To achieve this goal, we have developed an ABM that leverages a high-resolution synthetic database based on real-world data to compare the current car-based scenario with the lightweight autonomous system, analyzing its implications in terms of fleet sizing, service level, and environmental impacts. Additionally, we explore the impact of different vehicle configurations and operational strategies, enhancing the understanding of the system's performance under varying conditions. The modular design of our model enables its easy adaptability to other urban areas within the United States. Lastly, our model has been integrated into an interactive tool that allows stakeholders to explore the model's performance in real time, helping to understand the system's behavior and trade-offs in a more intuitive way. 

This research sheds light on the transformative potential of autonomous lightweight vehicles in the food delivery industry, offering valuable insights for stakeholders such as policymakers, mobility operators, and citizens. Through our evaluation of the performance of lightweight autonomous systems in urban food deliveries, informed decisions can be made regarding fleet design and implementation, with particular attention to environmental considerations. This study represents a crucial step towards understanding the capabilities and impacts of lightweight autonomous systems, paving the way for future advancements in the field.

\section{Modeling approach}
\label{sec:Modeling}

To study the performance of shared lightweight autonomous vehicles for food deliveries and compare their performance to the current car-based systems, we have developed an Agent-Based Model (ABM) and an environmental impact study which are detailed in Section \ref{sec:Architecture} and Section \ref{sec:EnvImpact}, respectively.

\subsection{Simulation model architecture}
\label{sec:Architecture}

ABMs have emerged as a popular method to analyze the meso- and macroscopic behavior of autonomous vehicles \citep{li2021systematic,  fagnant2014travel, sanchez2022simulation}. Their effectiveness lies in their ability to capture the complex interactions between various actors, such as vehicle fleets, users, and the infrastructure. Additionally, these models offer flexibility in exploring different scenarios and hypotheses, which is essential for understanding the uncertainties associated with emerging technologies.

\begin{figure}[!h]
    \centering
    \caption{Diagram that depicts the process of food delivery orders.}
    \includegraphics[width=0.8\linewidth]{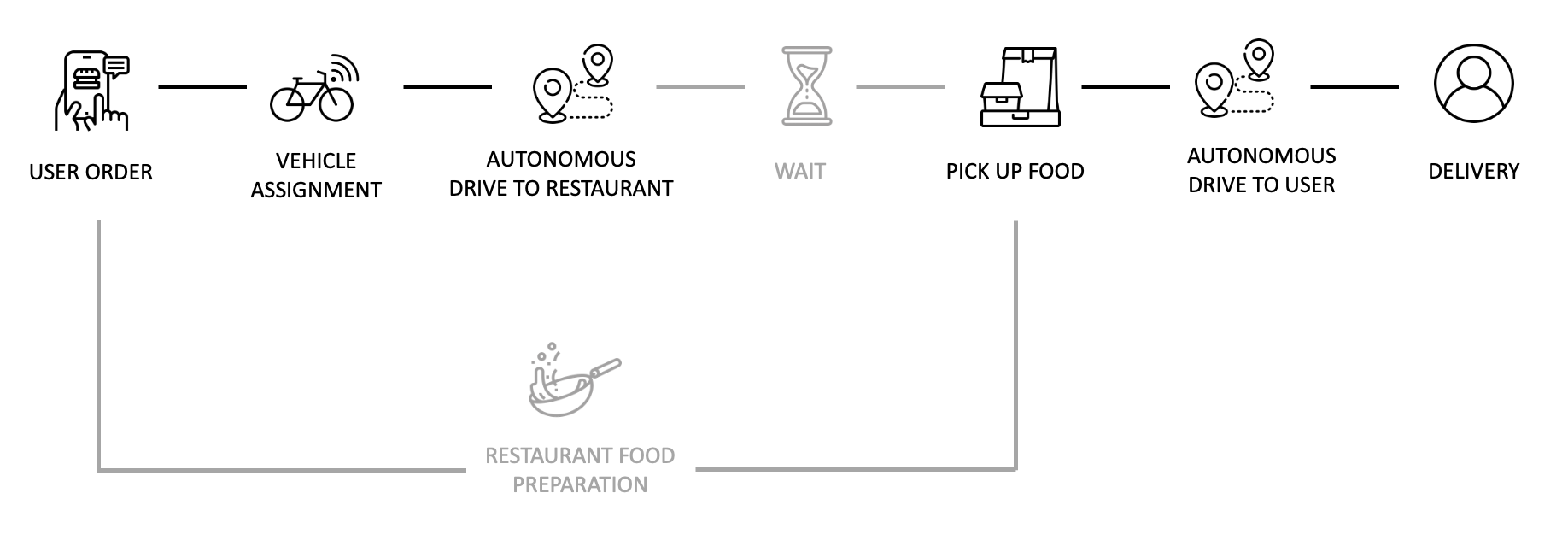}
    \label{fig:Activity Diagram}
\end{figure}

The ABM in this study is designed to capture the dynamics and interactions between the different agents involved in food delivery processes. Figure \ref{fig:Activity Diagram} provides a visual representation of the defined activities and their flow. First, the customer places a food delivery order in a particular restaurant. Subsequently, a vehicle is assigned to fulfill the delivery trip. The assigned vehicle drives to the restaurant and picks up the package. With the food onboard, the vehicle proceeds to the destination location where it will deliver the order.

\begin{figure}[!h]
    \centering
    \caption{Diagram for the depiction of the structure and interdependencies among the agent-based simulation layers: A) Urban infrastructure, B) Delivery vehicle fleet, C) User demand}
    \includegraphics[width=0.4\linewidth]{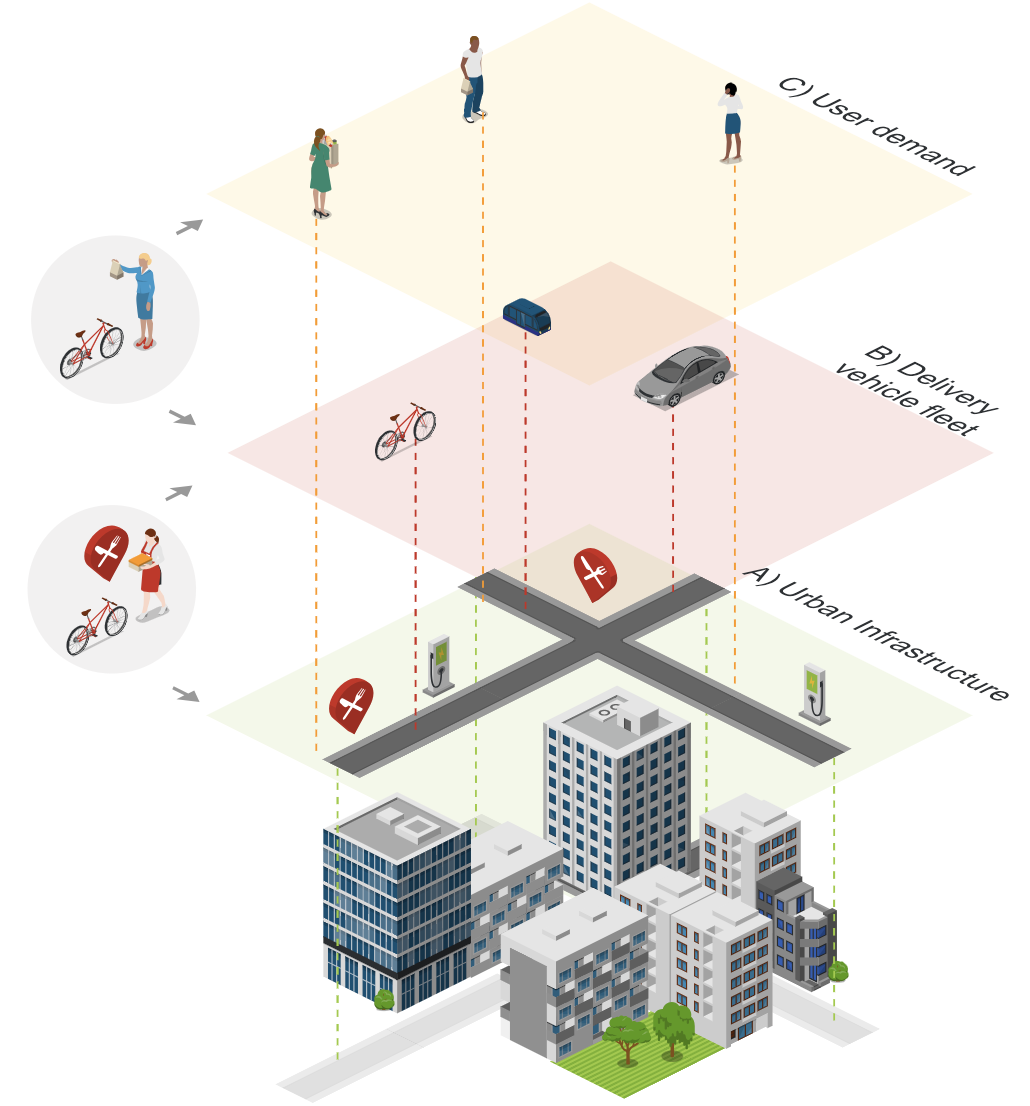}
    \label{fig:Simulation Layers Diagram}
\end{figure}

The ABM architecture consists of three interconnected layers, as shown in Figure \ref{fig:Simulation Layers Diagram}: A) the urban infrastructure, B) the delivery vehicle fleet, and C) the user demand. The following subsections will detail what these layers represent and how they have been modeled.

\subsubsection{Urban infrastructure}

The urban infrastructure plays a crucial role in shaping the operations and dynamics of the food delivery system. In this model, the urban infrastructure is represented by several components (Figure \ref{fig:Simulation Layers Diagram}-A). Firstly, the city road network represents the paths that vehicles will follow for their trips. It encompasses the network of streets and intersections within the area under study. Secondly, the buildings serve as the origins (i.e., restaurants) and destinations of the food delivery trips. Thirdly, the currently existing gas and charging stations represent the locations where vehicles will refuel or recharge their batteries. More details on the specific datasets used will be provided in Section \ref{sec:Experimental setup}.

\subsubsection{Vehicle behavior}
The ABM considers different scenarios with distinct fleets of vehicles to fulfill food deliveries (Figure \ref{fig:Simulation Layers Diagram}-B). In the baseline scenario, conventional cars are used, while the rest of the scenarios model a fleet of shared lightweight autonomous vehicles. The behaviors of these vehicle systems are defined as follows:

\begin{itemize}
    \item Baseline scenario: \textit{Current car-based deliveries.} This scenario models combustion cars to represent current food deliveries and electric cars to represent a futuristic but closer-to-date evolution of such deliveries. The behavior of these vehicles is modeled as a Finite State Machine (FSM), illustrated in red in Figure \ref{fig:Combined Vehicle Diagram}. 
    
    \begin{figure}[!htb]
        \centering
        \captionsetup{width=0.75\linewidth}
        \caption{Diagram of the Finite State Machine (FSM), representing the behavior and transitions between states of the cars (in red), which are modeled as part of the baseline scenario, and shared lightweight autonomous vehicles (in green), as part of the future scenario.}
        \includegraphics[width=0.7\linewidth]{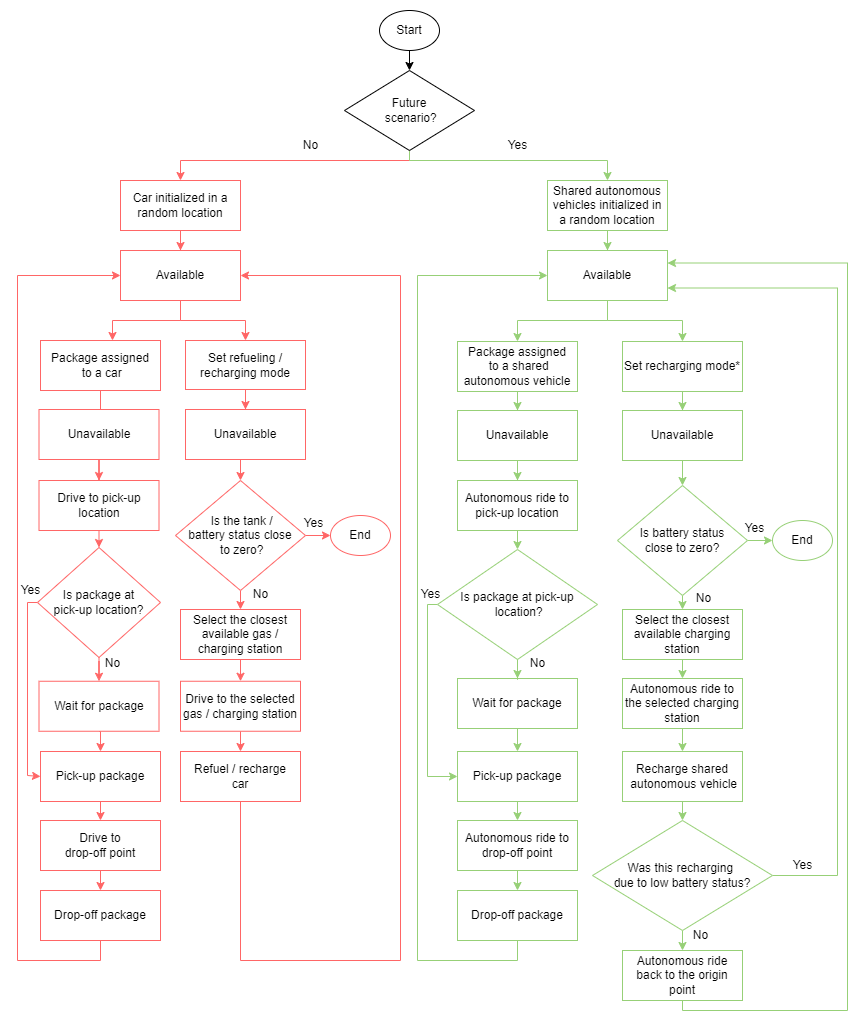}
        \label{fig:Combined Vehicle Diagram}
    \end{figure}

    Initially, cars are randomly placed on the city roads with a fuel/battery level set between the maximum and minimum values. All vehicles start in an available state, ready to respond to any food delivery request. When a customer orders at a specific restaurant, a package pick-up request is sent to the nearest available car with sufficient fuel/charge to complete the trip. The car then travels to the restaurant to collect the food order and delivers it to the designated drop-off location. After completing the delivery, the vehicle becomes available again and idles until it gets assigned a new pick-up request. If the fuel/battery level is low, it drives to the closest gas/charging station for refueling/recharging. Once refueled/recharged, the vehicle becomes available for further deliveries.

    \item Future scenario: \textit{Shared lightweight autonomous vehicles.} The behavior of shared lightweight autonomous vehicles is also modeled as a FSM, as depicted in green in Figure \ref{fig:Combined Vehicle Diagram}. The FSM captures various operational states of the vehicles, such as idle, in route, and delivering, and describes the transitions between these states.
    
    Shared lightweight autonomous vehicles are initialized at random locations within the road network, with an arbitrary battery level between the minimum and maximum thresholds. All vehicles start in an available state for food delivery trips. When an order is placed, a package pick-up request is assigned to the nearest available vehicle or the vehicle with the best distance-to-battery-level ratio, depending on the charging strategy being analyzed. The chosen vehicle autonomously travels to the restaurant, collects the package, and drives to the drop-off point. After completing the delivery, the vehicle becomes available again and idles until a new order is assigned. If the battery level falls below the minimum threshold, it autonomously drives to the closest available charging station, recharges, and becomes available again. Since different charging operational strategies have been studied (Section \ref{sec:Scenario definition}), the conditions under which vehicles initiate a recharge trip vary depending on the specific strategy being analyzed.

\end{itemize}

\subsubsection{Customer behavior}

\begin{figure}[!h]
    \centering
    \captionsetup{width=0.75\linewidth}
    \caption{Diagram of the Finite State Machine (FSM), representing the behavior and transitions between states of the food delivery orders that have been placed by the consumers.}
    \includegraphics[width=0.7\linewidth]{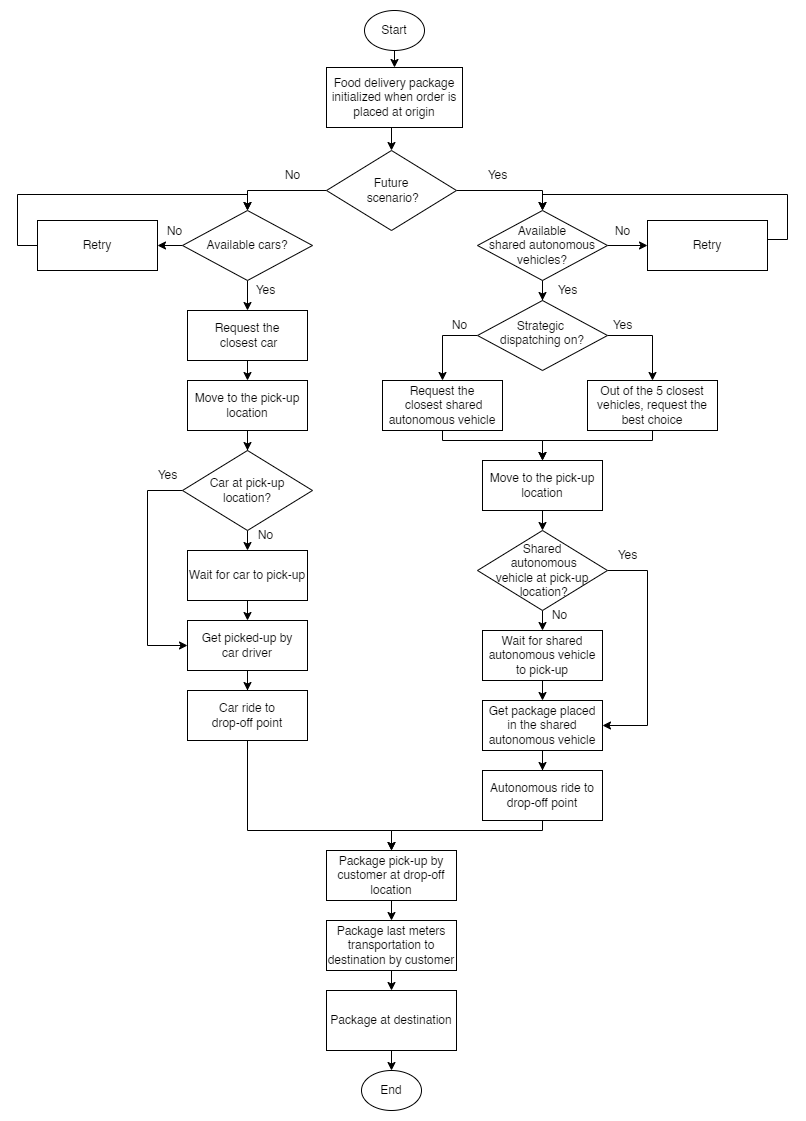}
    \label{fig:Food Delivery Package Diagram}
\end{figure}

This layer represents the customers who place food delivery orders (Figure \ref{fig:Simulation Layers Diagram}-C). The customers' behavior is also modeled as a FSM, which is illustrated in Figure \ref{fig:Food Delivery Package Diagram}. Whenever a user places a food delivery order at a specific restaurant, a food delivery package is generated at that location. As a first step, the system checks for the availability of vehicles. If no vehicles are available, the package will continuously attempt to find an available vehicle. If multiple vehicles are available, the system will determine which vehicle to assign to that delivery based on either the closest distance or the one with the best proximity-to-battery ratio, as defined in Section \ref{sec:Scenario definition}. Once a delivery vehicle is assigned, the package will be transported to its designated delivery location, where the customer will receive it.

\subsection{Environmental impact modeling}\label{sec:EnvImpact}

While the ABM in Section \ref{sec:Architecture} aims to analyze the systems from a performance perspective, understanding the corresponding environmental impacts is key in analyzing the potential implications of shared lightweight autonomous vehicles for food deliveries.

The environmental impacts considered for this study are based on a life cycle assessment (LCA). LCA is a standardized environmental impact calculation method that considers the impacts from the entire life cycle: from raw material extraction to production, use, and waste management \citep{ISO2006,finnveden2009recent}. This study focuses on the impacts in terms of $\mathrm{CO_2}$ emissions because it is a metric that plays a central role in the transportation and governmental decision-making processes \citep{EPA2021}.

The environmental assessments in this study draw upon the methodologies outlined in \citep{sanchez2022can}, which have been suitably adapted to this specific case study. These adaptations involve the customization of vehicle utilization rates and battery ranges, guided by simulation outcomes. Notably, infrastructure-related impacts are intentionally omitted from the analysis because all scenarios consider a consistent number of charging stations. This approach is adopted to prevent potential bias introduced by not sizing charging infrastructure proportionally to the number of vehicles in operation.

Our assessment of the environmental impacts unfolds through several key steps. Firstly, the simulation results in \ref{subsec:Fleet level performance} provide data on average kilometers covered by both vehicle types (cars and shared lightweight autonomous vehicles) across various vehicle configurations (speeds and battery ranges). Second, leveraging the environmental impacts calculation process presented in \citep{sanchez2022can}, we have calculated the grams of $\mathrm{CO_2}$ per kilometer traveled. Lastly, for the purpose of comparing environmental impact reductions between scenarios, the total distance traveled by each system and the associated grams of $\mathrm{CO_2}$ per kilometer traveled are taken into consideration.  In order to model the fast-charging method, twice as many batteries per vehicle have been considered to account for the battery-swapping process.

\section{Experimental setup}
\label{sec:Experimental setup}

\subsection{Agent-based modeling software}
\label{subsec:Modeling software}

The ABM in this study has been developed using the GAMA Platform \citep{taillandier2019building, grignard2013gama}. GAMA is an open-source tool specifically designed for spatially explicit multi-layer agent-based simulations. It has successfully been employed in various domains, including  urban decision-making tools \citep{alonso2018cityscope}, epidemiology representation \citep{drogoul2020designing,palomo2022agent}, and social simulation \citep{caillou2017simple,ayoub2022proxymix}.

\subsection{Case study location} 
While the model can be easily applied to any US city, we have chosen Cambridge, MA, as the case study for this research. Cambridge has a population of approximately 118,403 residents and covers an area of 6.39 square miles \citep{census2020}. The boundary of the city can be observed in Figure \ref{fig:Food Delivery Demand}.

The road network and buildings information have been integrated in the model in the form of GIS data, extracted from \citet{OSM}, and the Cambridge, MA, government website \citep{cambridgemaGIS}.  The data regarding the gas and charging station locations has been collected from \citet{OSM} and \citet{bluebikes} data, and also provides detailed information including their current capacity and location.

\subsection{Input demand dataset} \label{sec:demand}

\begin{figure}[!h]
    \centering
    \caption{Simplified diagram of the synthetic database generation process for obtaining the fine-grained food delivery demand dataset.}
    \includegraphics[width=0.9\linewidth]{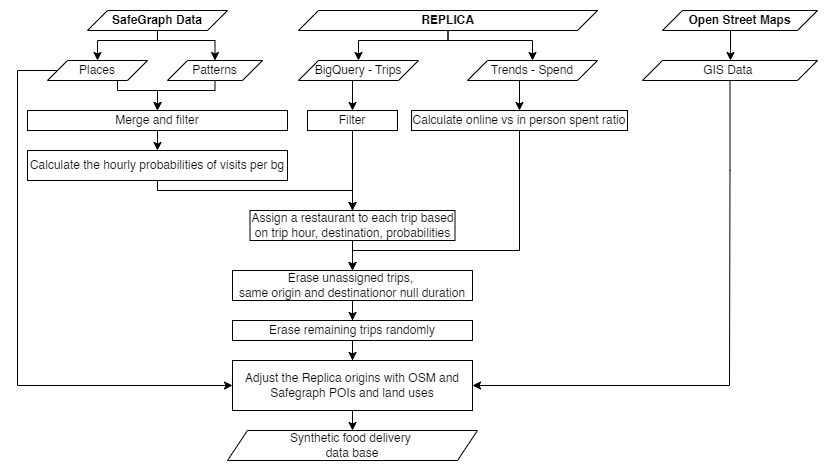}
    \label{fig:Simplified Synthetic Database Generation}
\end{figure}

Since open-source data on food deliveries is limited, we have generated a synthetic demand dataset, based on three different sources. Firstly, general land use data was obtained from  \citet{OSM}. Secondly, data on trips made to go to restaurants and bars was obtained through \citet{replica}, which also provided a breakdown of online versus in-person food expenditures. Lastly, data on the popularity of specific restaurants within each blockgroup was obtained through \citet{safegraph}.  A simplified outline of methodology for generating the demand dataset is depicted in Figure \ref{fig:Simplified Synthetic Database Generation}. For more detailed information, please refer to Appendix \ref{app:Tables and Figures}, Figure \ref{fig:Synthetic Database Generation}.

 The first step was to filter Replica trips with origin and destination in Cambridge, and travel purpose of eating. Since Replica provides origins and destinations at block-group level, we assigned a specific restaurant from SafeGraph to each trip based on the popularity and open hours of the restaurants. To assign the origin of the trips, we created two association tables (Appendix \ref{app:Tables and Figures} Figure \ref{tab:Land use association Replica} and Figure \ref{tab:Land use association SafeGraph}) that link the building land use types defined in Replica with the land uses provided by SafeGraph and Open Street Map. This allowed us to assign an exact location to each trip based on the land use of the origin building.

After assigning specific origin and destinations to each trip, the dataset was filtered to eliminate any unreasonable trips: we dropped all trips without a restaurant assignment, trips without duration, and trips with the same coordinates for origin and destination. Then, the number of trips was scaled proportionally to in-person versus online expenditures reported by \citet{replica} and their average order size \citep{kimes2014customer,npd2020,1010data2016,Coppola2022statista,ahuja2021ordering}. Since there are fewer online orders than in-person orders, the extra trips were removed randomly. As a last step, the origin and destination of the trips were inverted to represent how food delivery trips work (from the restaurant to any other location).

\begin{figure}[!h]

    \includegraphics[width=0.7\linewidth]{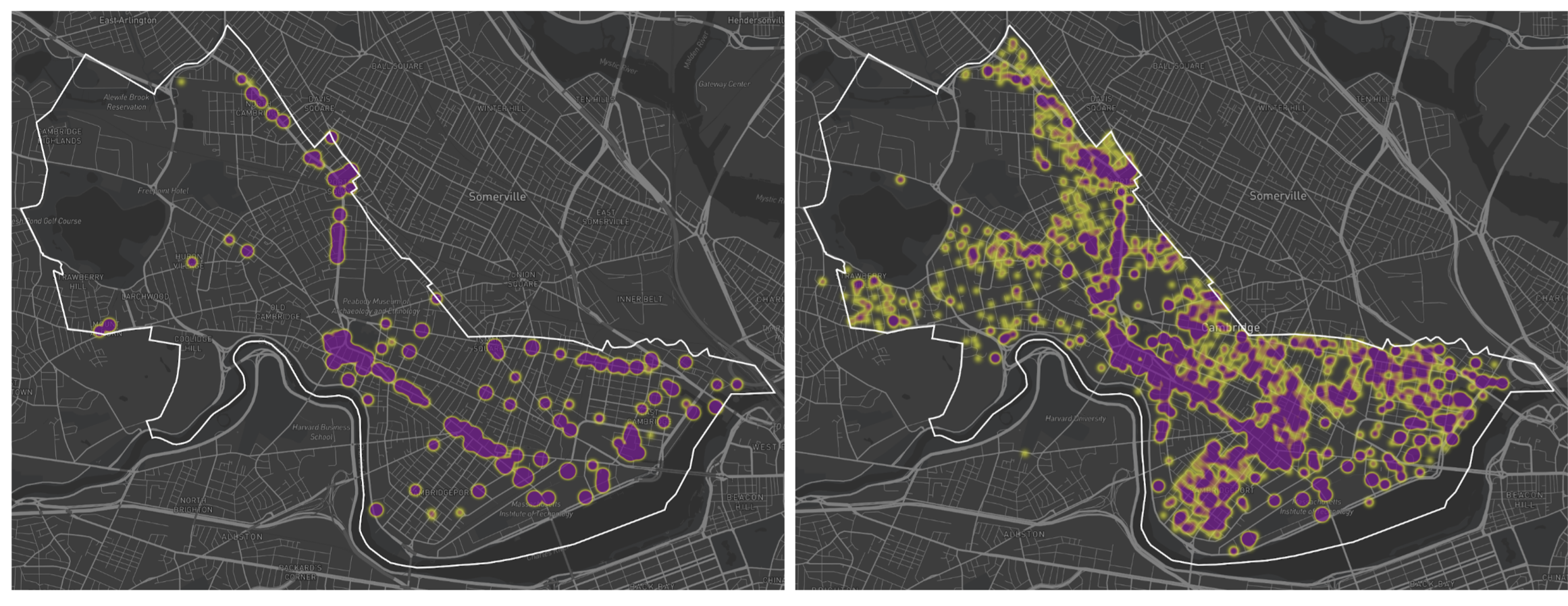}
    \includegraphics[width=0.5\linewidth]{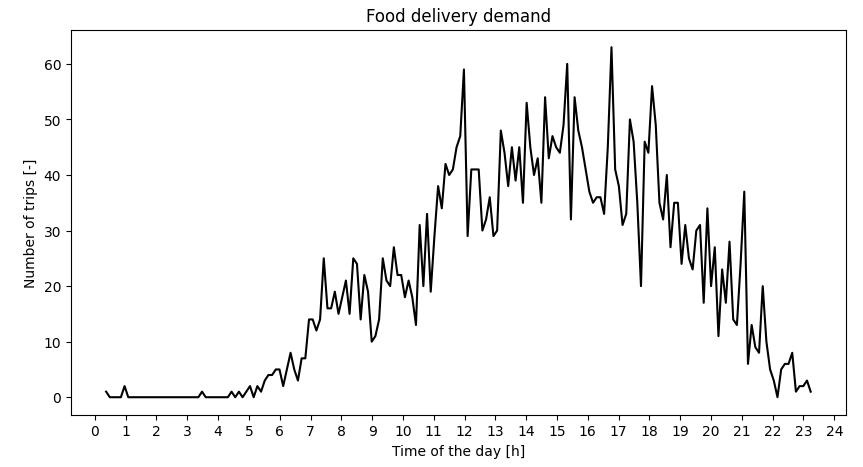}
    \caption{Top: Heat map illustrating the spatial density of trip origins (left, restaurants) and destinations (right), with areas of highest density represented in violet. The map also shows the boundary of the study area, which encompasses Cambridge, MA, USA. Bottom: Demand profile of food delivery orders in the study area (Cambridge, MA, USA) demonstrating the temporal distribution of orders throughout the day, aggregated by time intervals of 7.5 minutes. }
    \label{fig:Food Delivery Demand}
\end{figure}

Using this methodology, we generated the food delivery demand dataset, consisting of 3,989 trips. Figure \ref{fig:Food Delivery Demand} illustrates heat maps of the spatial density of trip origins and destinations and the number of food delivery orders placed at different times of the day.

\subsection{Scenario definition}
\label{sec:Scenario definition}

In order to understand and analyze the performance of shared lightweight autonomous vehicles for food deliveries, we compared them to a baseline scenario that represents the way food deliveries are done now.  

The baseline scenario is composed of two different sub-scenarios: The first sub-scenario models food deliveries using internal combustion engine cars (ICE) as a baseline, while the second sub-scenario models battery electric vehicles (BEV) to account for the ongoing transition towards them.

In addition to the baseline scenario, a lightweight autonomous vehicle-based future scenario has also been studied through several operational decisions. Various system design parameters can be defined when designing a new mobility service to achieve the desired performance. Hence, we have examined the potential implications of different choices for these design parameters, including different vehicle configurations (i.e., battery sizes and autonomous driving speeds) and operational strategies related to vehicle charging (i.e., conventional, fast, and night charging, and strategic dispatching). The sub-scenarios considered in the lightweight autonomous vehicle scenario are the following:

\begin{itemize}
    \item Conventional charging (CC): This sub-scenario represents vehicles being charged at conventional plug-in charging stations, taking 4.5 hours for a full-battery recharge, which reflects the performance of current lightweight electric vehicle charging processes \citep{ITF2020}.
    
    \item Night charging (NC): In this sub-scenario, charging stations are conventional charging stations like in the CC scenario. However, vehicles with less than 90\% battery charge get recharged during the night, coinciding with the lowest demand period (2-5 am).
    
    \item Strategic dispatching (SD): This sub-scenario builds upon the NC scenario by adding a strategic condition for dispatching. In such strategic dispatching, instead of assigning the nearest available vehicle to each food delivery order, the dispatcher considers up to five nearest available vehicles and then assigns the one with the manually adjusted best distance to battery level ratio. 
    
    \item Fast charging (FC): In this sub-scenario, stations are battery swapping stations instead of plug-in stations. Battery swapping is a process where depleted batteries are replaced with fully charged ones. This process eliminates the need to wait for the battery to recharge. Therefore, the charging process is considered to take 1.85 min, which is an average of the two battery-swapping scenarios studied in \citep{huang2020understanding}.
    
\end{itemize}

A comprehensive summary of the scenarios analyzed and their parameters can be found in Table \ref{tab:Scenarios & parameters}.

\begin{table}[!h]
    \centering
    \caption{Overview of the scenarios examined in the study, including detailed specifications of parameter values that define their operational characteristics and behavior.}
    \resizebox{\textwidth}{!}{%
        \begin{tabular}{ccccccc}
            \hline
            Scenario & Nomenclature & Charging Technology & Full Recharge Time & Minimum Battery Level & Riding Speeds {[}km/h{]} & Battery Autonomy {[}km{]} \\ \hline
            Baseline (Cars) & ICE & Combustion            & 3 minutes          & 15\%  & 30          & 500                       \\
            Baseline (Cars) & BEV & Electric              & 30 minutes         & 15\%  & 30          & 342                       \\
            Future (SLAV)    & CC  & Conventional Charging & 4 hours 30 minutes & 25\%  & 8 - 11 - 14 & 35 - 50 - 65              \\
            Future (SLAV)    & NC & Night Charging         & 4 hours 30 minutes & 25\%  & 8 - 11 - 14 & 35 - 50 - 65              \\
            Future (SLAV)    & SD & Strategic Dispatching  & 4 hours 30 minutes & 25\%  & 8 - 11 - 14 & 35 - 50 - 65              \\
            Future (SLAV)    & FC & Fast Charging          & 1.85 minutes       & 25\%  & 8 - 11 - 14 & 35 - 50 - 65              \\ \hline
        \end{tabular}%
    }
    \label{tab:Scenarios & parameters}
\end{table}

\subsection{Vehicle modeling}

The analysis of the different scenarios and sub-scenarios described in Subsection \ref{sec:Scenario definition} required different vehicles to be modeled in the ABM. In this section, we define the modeling assumptions for each vehicle type, which are also summarized on Table \ref{tab:Scenarios & parameters}. 

In the baseline scenario, ICE cars are considered to have a driving autonomy of 500 km and a refueling rate of 3 minutes \citep{straus2022, wishartfuel}. For BEV cars, instead, the study assumes an autonomy of 342 km, which is the average range reported by \citet{EVDatabase2023} and a recharging rate of 30 minutes \citep{Sanguesa2021ARO}. In both cases, cars have been modeled to travel at a constant speed of 30 km/h \citep{cambridgemagov2022,replica}. In addition, they are also considered to transition into the refuel/recharging state when their low gas/charge level is below 15\% of the total tank/battery \citep{aceable2022}. Gas and charging stations are located at the same locations as current gas stations in Cambridge \citep{OSM} and considered to have the same capacity as them.

In the future scenario that models shared lightweight autonomous vehicles, instead of modeling their behavior with fixed parameters, we have analyzed different operational decisions that include several values proposed in \citet{sanchez2022performance}. This is due to the fact that modeling this emerging technology holds uncertainties regarding its real-world performance and, as indicated by previous studies, vehicle configuration parameters and charging operational strategies have a direct and very significant influence on fleet-level performance \citep{chen2016operations,sanchez2022simulation,sanchez2022performance}. As a consequence, we have modeled several battery sizes (35-50-65 km), driving speeds (8-11-14 km/h), and charging strategies defined in Section \ref{sec:Scenario definition}. Shared lightweight autonomous vehicles are considered to enter a low battery state when their battery level is bellow 25\% of their total capacity, and charging stations have been considered to be at the same locations and have the same capacities as the current \citet{bluebikes} docking stations. The charging threshold has been considered to be higher than it is for cars because autonomous vehicle operators need to be more conservative in ensuring that vehicles never run out of battery before reaching a charging station.

\section{Results and discussion}
\label{sec:Results and discussion}

This section presents the simulations' results aimed at evaluating performance of shared lightweight autonomous systems for food deliveries. As discussed in Section \ref{sec:Scenario definition}, we compare the performance of this new system with the current car-based delivery system, considering both internal combustion engine (ICE) cars and battery electric vehicles (BEV). Moreover, we analyze different vehicle configurations and operational strategies in lightweight autonomous systems. The summary of all the scenarios and sub-scenarios considered can be found in Table \ref{tab:Scenarios & parameters}. 

To facilitate a fair comparison, all systems in the study adhere to the same design criteria. In line with previous studies \citep{sanchez2022performance,chen2016operations}, this criteria has been based on a desired service level. Specifically, we have defined a quality standard requiring all trips to be served, with 95\% of the trips taking less than 40 minutes from order to delivery, based on a national survey data by \citet{USFoods2019}.

The assessment of the lightweight autonomous system's performance has been approached from two distinct angles. Firstly, in Section \ref{subsec:Fleet level performance}, an in-depth exploration of fleet-level performance is conducted, analyzing the interplay of various scenarios and parameters on the overall system performance. Secondly, in Section \ref{subsec:Environmental impacts}, these findings are leveraged to evaluate the corresponding environmental impacts. This comprehensive approach not only sheds light on aspects like user wait times and fleet sizes, but also delves into the environmental consequences of this potential transition to shared lightweight autonomous vehicles.

\subsection{Fleet level performance}
\label{subsec:Fleet level performance}

This section evaluates the fleet-level performance across the different scenarios and sub-scenarios considered in this study. The exploration encompasses a detailed analysis of diverse vehicle configurations and charging operational strategies. 

We first analyze the baseline scenario, which represents the current car-based deliveries in Section \ref{subsubsec:Baseline scenario: Cars}. Subsequently, we analyze the fleet-level performance of lightweight autonomous systems, as elaborated in Section \ref{subsec:Future scenario: Shared lightweight autonomous}.

\subsubsection{Baseline scenario: Current car-based deliveries}
\label{subsubsec:Baseline scenario: Cars}

\begin{figure}[!htb]
      \includegraphics[width=0.8\textwidth]{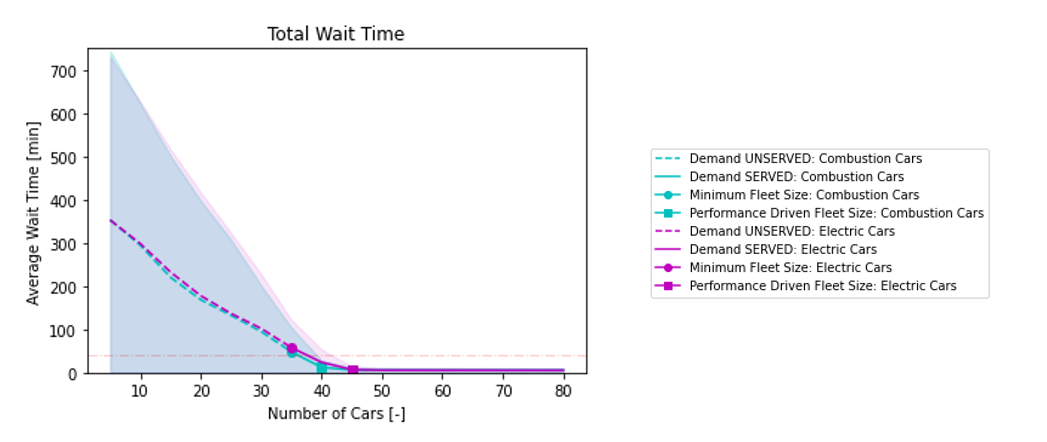}
          \caption{Variation of the average wait time for food delivery orders with an increasing number of electric and combustion engine cars. The dashed horizontal line represents the desired service level requirement.}%
          \label{fig:Average wait time vs fleet size: cars}
\end{figure}

This scenario's main objective, described in Section \ref{sec:Scenario definition}, is to provide a baseline scenario against which we can evaluate shared lightweight autonomous systems. We present the main results obtained from the simulation in Figure \ref{fig:Average wait time vs fleet size: cars} and Table \ref{tab:Baseline scenario results}. 

\begin{table}[!htb]
    \centering
    \resizebox{0.4\textwidth}{!}{%
       \begin{tabular}{lcc}
        \hline
        \multicolumn{3}{c}{\textbf{Baseline scenario results}} \\ \hline
        \multicolumn{1}{c}{Metric} & {ICE} & {BEV} \\  \hline
        Num. of cars {[}-{]} & 40 & 45                                    \\
        Food delivery demand {[}-{]} & 3989 & 3989                  \\
        Avg. trip time {[}min{]} & 12.71 & 6.19                            \\
        Trips under 40 min {[}\%{]} & 99.82\% & 100\%                     \\
        Avg. trips/car/day {[}-{]} & 99.72 & 88.64                        \\
        Total refuelings/day {[}-{]} & 33 & 37                            \\
        Total vehicle km {[}km{]} & 10089.48 & 8681.82             \\
        | Avg. km for pick-up {[}\%{]} & 54.03\% & 46.58\%        \\
        | Avg. km for delivery {[}\%{]} & 45.66\% & 53.06\%       \\
        | Avg. km to refill {[}\%{]} & 0.31\% & 0.36\%             \\
        Avg. km/car {[}km/car{]} & 252.24 & 192.93                 \\ \hline
        \end{tabular}
    }
      \caption{Summary of the main performance metrics in the baseline scenario, which models the way food deliveries are currently done by ICE vehicles, as well as BEV. }%
      \label{tab:Baseline scenario results}%

\end{table}

As can be observed in Table \ref{tab:Baseline scenario results}, the specified service requirements can be met by serving the total demand of 3,989 trips with 40 ICE cars. Figure \ref{fig:Average wait time vs fleet size: cars} shows how the average wait time decreases with increased fleet sizes until the service level requirement is met. The main impact that can be expected from the transition towards BEVs from the fleet performance perspective is a slight increase in the required fleet size. As shown in Table \ref{tab:Baseline scenario results}, the same demand can be served with 45 BEVs. This is due to the longer recharge time for BEVs than the refueling time for ICE cars. However, it is noteworthy that the overall behavior and the relationship between service level and fleet size remain similar in both cases.

\subsubsection{Future scenario: Shared lightweight autonomous vehicles}
\label{subsec:Future scenario: Shared lightweight autonomous}

This scenario presents the results of the fleet performance of shared lightweight autonomous vehicle-based food deliveries, considering different vehicle configurations and charging operational strategies.

\begin{itemize}
    \item \textit{Vehicle configurations} 
    \label{subsubsec:Vehicle configurations}
    \\[0.2cm]
   The variations in service level under different vehicle configurations and fleet sizes are summarized in Table \ref{tab:Future scenario results} and illustrated in Figure \ref{fig:avgwaits}. Table \ref{tab:Future scenario results} demonstrates that the fleet size required to meet the demand with the desired service level ranges from 170 to 310, depending on the chosen vehicle configurations. As anticipated, these fleet sizes are notably larger than the corresponding car fleets due to the lower speeds of vehicles (8-14 km/h versus 30 km/h). However, as discussed in Section \ref{subsec:Environmental impacts}, despite the increased fleet sizes, the lightweight nature of each vehicle offers potential improvements in environmental impacts.

\begin{figure}[!h]
    \centering
    \includegraphics[width=0.9\linewidth]{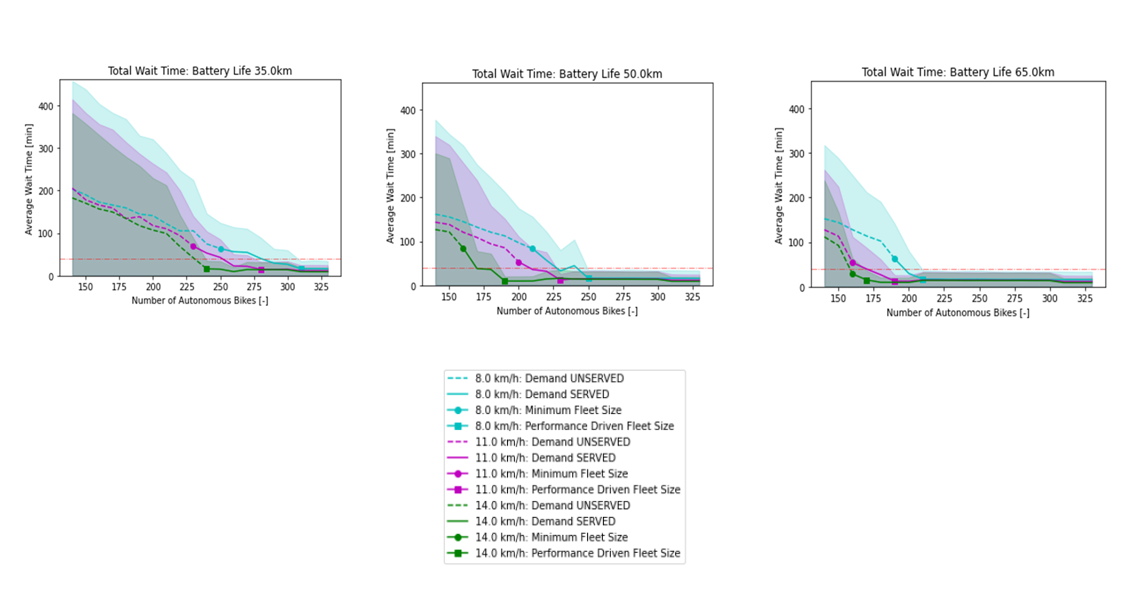}
    \caption{Variation on the average wait time for food delivery orders for different fleet sizes. Each figure represents a certain battery capacity (35 km, 50 km, and 65 km, left to right).}
    \label{fig:avgwaits}
\end{figure}

    \begin{table}[!h]
        \centering
        \resizebox{\textwidth}{!}{%
            \begin{tabular}{|lccccccccc|}
                \hline
                \multicolumn{10}{|c|}{  \textbf{Future scenario (SLAV) results}} \\ \hline
                \multicolumn{1}{|c|}{Battery size} & \multicolumn{3}{c|}{Small} & \multicolumn{3}{c|}{Medium} & \multicolumn{3}{c|}{Large}   \\ \hline
                \multicolumn{1}{|c|}{Speed} & Slow & Medium & \multicolumn{1}{c|}{Fast} & Slow & Medium & \multicolumn{1}{c|}{Fast} & Slow    & Medium & Fast    \\ \hline
                \multicolumn{1}{|l|}{Num. of SLAV {[}-{]}} & 310 & 280 & \multicolumn{1}{c|}{240} & 250 & 230 & \multicolumn{1}{c|}{190} & 210 & 190 & 170 \\
                \multicolumn{1}{|l|}{Food delivery demand {[}-{]}} & 3989 & 3989 & \multicolumn{1}{c|}{3989} & 3989 & 3989 & \multicolumn{1}{c|}{3989} & 3989 & 3989 & 3989    \\
                \multicolumn{1}{|l|}{Avg. trip time {[}min{]}} & 15.9 & 14.94 & \multicolumn{1}{c|}{15.93} & 16.39 & 12.09 & \multicolumn{1}{c|}{9.67} & 16.98 & 12.34 & 15.2    \\
                \multicolumn{1}{|l|}{Trips under 40 min {[}\%{]}} & 97.77\% & 98.09\% & \multicolumn{1}{c|}{97.72\%} & 97.39\% & 99.72\% & \multicolumn{1}{c|}{100.00\%} & 96.14\% & 99.65\% & 97.27\% \\
                \multicolumn{1}{|l|}{Avg. trips/SLAV/day {[}-{]}} & 12.8 & 14.25 & \multicolumn{1}{c|}{16.62} & 15.96 & 17.34 & \multicolumn{1}{c|}{20.99} & 19 & 20.99 & 23.46   \\
                \multicolumn{1}{|l|}{Total charges/day {[}-{]}} & 317 & 351 & \multicolumn{1}{c|}{449} & 261 & 239 & \multicolumn{1}{c|}{244} & 202 & 198 & 242     \\
                \multicolumn{1}{|l|}{Total vehicle km {[}km{]}} & 6695.45 & 6935.2 & \multicolumn{1}{c|}{9209.01} & 6919.22 & 6892.96 & \multicolumn{1}{c|}{6899.58} & 7178.07 & 7046.08 & 8902.37 \\
                \multicolumn{1}{|l|}{| Average km for pick-up {[}\%{]}} & 30.30\% & 32.60\% & \multicolumn{1}{c|}{49.02\%} & 32.69\% & 32.47\% & \multicolumn{1}{c|}{32.55\%}  & 35.24\% & 34.05\% & 47.68\% \\
                \multicolumn{1}{|l|}{| Average km for delivery {[}\%{]}} & 68.80\% & 66.42\% & \multicolumn{1}{c|}{50.02\%} & 66.58\% & 66.83\% & \multicolumn{1}{c|}{66.77\%}  & 64.18\% & 65.38\% & 51.75\% \\
                \multicolumn{1}{|l|}{| Average km to recharge {[}\%{]}} & 0.89\%  & 0.97\%  & \multicolumn{1}{c|}{0.95\%}  & 0.74\%  & 0.70\%  & \multicolumn{1}{c|}{0.68\%}   & 0.58\%  & 0.57\%  & 0.57\%  \\
                \multicolumn{1}{|l|}{Average km/SLAV {[}km/SLAV{]}} & 21.6 & 24.77 & \multicolumn{1}{c|}{38.37} & 27.68 & 29.97 & \multicolumn{1}{c|}{36.31} & 34.18 & 37.08 & 52.37   \\ \hline
            \end{tabular}%
        }
            \caption{Summary of the main performance metrics in a food delivery system based on a fleet of shared  lightweight autonomous vehicles. Different sub-scenarios account for different battery sizes and autonomous driving speeds.}%
            \label{tab:Future scenario results}
    \end{table}

    The results in Table \ref{tab:Future scenario results} and Figure \ref{fig:avgwaits} highlight the significant dependence of vehicle configurations on the required fleet sizes. The largest fleet size is almost twice as big as the smallest one, due to a number of factors: Firstly, faster vehicles can complete tasks quicker, reducing the number of vehicles needed. Additionally, larger battery ranges minimize the frequency of charging trips, resulting in greater vehicle availability and smaller required fleet size.
    
    Notably, the transition from small to medium batteries has a more pronounced effect on fleet size improvement than the transition from medium to large batteries. Similarly, the increase in autonomous driving speed has a more substantial impact when transitioning from medium to high speed compared to the transition from low to medium speeds. These findings indicate that fleet operators can make specific cost-benefit trade-offs when deciding on vehicle configurations.

    \item \textit{Charging operational strategies}
    \label{subsubsec:Charging operational strategies}
    \\[0.2cm]

    This section examines different charging strategies and their potential impact on system performance and fleet size requirements. As discussed in Section \ref{sec:Scenario definition}, four charging strategies were studied: conventional charging (CC), night charging (NC), strategic dispatching (SD), and fast charging (FC). The fleet size variation in each  scenario is presented in Table \ref{tab:Fleet size variation with different charging operational strategies}. This table reveals that the NC strategy can reduce the fleet size needed for small and medium battery sizes and low to medium travel speeds, but it has a negative impact at high speeds. Similarly, the SD strategy only shows a positive impact at slower speeds. 
    
    On the other hand, the FC strategy consistently and significantly reduces the required fleet size across all scenarios, with reductions ranging from 28.57\% to 62.5\%. Notably, the impact is more pronounced for smaller battery sizes due to their higher reliance on charging events. In fact, in the FC scenario, the minimum fleet size needed to meet the quality standard is independent of the battery size.

    \begin{table}[!h]
        \centering
        \resizebox{0.9\textwidth}{!}{%
            \begin{tabular}{|c|c|c|cc|cc|cc|}
                \hline
                Battery & Speed & Conventional Charging (CC) & \multicolumn{2}{c|}{Night Charging (NC)} & \multicolumn{2}{c|}{Strategic Dispatching (SD)} & \multicolumn{2}{c|}{Fast Charging (FC)} \\ \hline
                & Slow & 310 & 260 & -16.13\% & 300 & -3.23\% & 150 & -51.61\%                         \\ 
                & Medium & 280 & 260 & -7.14\% & 280 & 0.00\% & 110 & -60.71\%                         \\ 
                \multirow{-3}{*}{Small} & Fast & 240 & 240 & 0.00\% & 270 & 12.50\% & 90 & -62.50\%      \\ \hline
                & Slow & 250 & 240 & -4.00\% & 240 & -4.00\% & 140 & -44.00\%                         \\ 
                & Medium & 230 & 210 & -8.70\% & 230 & 0.00\% & 110 & -52.17\%                         \\ 
                \multirow{-3}{*}{Medium} & Fast & 190 & 210 & 10.53\% & 230 & 21.05\% & 90 & -52.63\%  \\ \hline
                 & Slow & 210 & 200 & -4.76\% & 180 & -14.29\% & 150 & -28.57\%                         \\ 
                & Medium & 190 & 190 & 0.00\% & 180 & -5.26\% & 110 & -42.11\%                          \\ 
                \multirow{-3}{*}{Large}  & Fast & 170 & 180 & 5.88\% & 180 & 5.88\% & 90 & -47.06\%           \\ \hline
            \end{tabular}%
            \caption{Fleet size needed to meet the service requirements under different operational strategies related to charging. The minimum required fleet size for each operational strategy is calculated for sub-scenarios with varying battery sizes and autonomous driving speeds.}%
            \label{tab:Fleet size variation with different charging operational strategies}
        }
    \end{table}

  For further insights to understand the reasons behind these phenomena refer to Figures \ref{fig:Vehicles state CC}, \ref{fig:Vehicles state NC}, \ref{fig:Vehicles state SD}, and \ref{fig:Vehicles state FC} in Appendix \ref{app:Tables and Figures}. These figures demonstrate that in CC, NC, and SD scenarios, the fleet size is strongly influenced by the dynamics of vehicle charging. However, in the FC scenario, this dependency is significantly reduced because the battery-swapping process is fast. 
  
  Figures \ref{fig:Vehicles state CC} and \ref{fig:Vehicles state NC} also illustrate that the peak of the concurrent number of vehicles charging in the NC scenario is comparable to that of the CC scenario. The peak is even higher in the SD scenario (Figure \ref{fig:Vehicles state SD}), while no peak is observed in the FC scenario (Figure \ref{fig:Vehicles state FC}). These results can be attributed to two synergistic effects: Firstly, charging vehicles at night in the NC and SD scenarios homogenizes the battery level across the fleet. Consequently, when vehicles start to deplete their batteries later in the day, there is a sudden surge of vehicles reaching low battery levels within a short period, resulting in the observed abrupt increase in the number of vehicles charging. Secondly, the SD strategy homogenizes the average distance traveled by vehicles per day by selecting vehicles based on the best distance-to-battery level ratio for trip assignments. This further concentrates the charging needs of vehicles in a specific period, leading to a more pronounced peak in the charging events.
  

In conclusion, two key findings emerge. Firstly, in the three scenarios involving conventional charging, the fleet size is constrained by the charging events. In contrast, in the fast charging (FC) scenario, the primary constraint is the demand itself. This significantly impacts the required fleet size, with the fast charging scenario requiring roughly half the fleet size compared to the other three scenarios. Secondly, night charging (NC) and strategic dispatching (SD) strategies provide limited benefits in reducing the required fleet size due to the homogenization of battery levels and vehicle kilometers traveled. Consequently, a delayed and more abrupt peak of charging events still limits the fleet sizing.
    
\end{itemize}

\subsection{Environmental impacts}
\label{subsec:Environmental impacts}

This section presents an overview of the potential reductions in the equivalent $\mathrm{CO_2}$ emissions of a shared lightweight autonomous vehicle-based system with the different vehicle configurations and charging operational strategies analyzed in Section \ref{subsec:Future scenario: Shared lightweight autonomous} and how they compare to current car-based systems analyzed in Section \ref{subsubsec:Baseline scenario: Cars}. Moreover, since current trends point towards the electrification of car fleets and a decarbonization of the electricity grid, we have also considered a scenario in which all cars would be electric, and the grid would be zero-carbon. This allows us to understand whether lightweight autonomous vehicles could also be environmentally beneficial in this potential future. The results of this analysis have been consolidated in Table \ref{tab:Environmental impacts}. 

\begin{table}[!h]
    \resizebox{\textwidth}{!}{%
        \begin{tabular}{cccccccccc}
        \cline{7-10}
                                                                       & & & & & \multicolumn{1}{c|}{} & \multicolumn{2}{c|}{US Electricity mix} & \multicolumn{2}{c|}{100\% Renewable} \\ \cline{2-10} 
        \multicolumn{1}{c|}{} & Fleet size & Battery km & Speed km/h & Avg. distance km & \multicolumn{1}{c|}{Total system km} & g$\mathrm{CO_2}$/km & \multicolumn{1}{c|}{\% red. vs ICE} & g$\mathrm{CO_2}$/km & \multicolumn{1}{c|}{\% red. vs BEV} \\ \hline
        \multicolumn{1}{|c|}{Combustion   Cars (ICE)} & 40 & 500 & 30 & 252.24 & \multicolumn{1}{c|}{10090} & 161.97  & \multicolumn{1}{c|}{} & 161.97 & \multicolumn{1}{c|}{} \\ \hdashline
        \multicolumn{1}{|c|}{Electric   Cars (BEV)} & 45 & 342 & 30 & 192.93 & \multicolumn{1}{c|}{8682} & 107.53 & \multicolumn{1}{c|}{} & 53.85 & \multicolumn{1}{c|}{} \\ \hline
                                                                       & & & & & & & & &                                       \\ \hline
        \multicolumn{1}{|c|}{\multirow{9}{*}{Conventional Charging (CC)}} & 310 & 35 & 8 & 21.6 & \multicolumn{1}{c|}{6696} & 44.95 & \multicolumn{1}{c|}{-81.58\%} & 36.07 & \multicolumn{1}{c|}{-48.34\%}         \\
        \multicolumn{1}{|c|}{} & 280 & 35 & 11 & 24.77 & \multicolumn{1}{c|}{6936} & 41.08 & \multicolumn{1}{c|}{-82.57\%} & 32.20 & \multicolumn{1}{c|}{-52.23\%}         \\
        \multicolumn{1}{|c|}{} & 240 & 35 & 14 & 38.37 & \multicolumn{1}{c|}{9209} & 31.16 & \multicolumn{1}{c|}{-82.44\%} & 22.27 & \multicolumn{1}{c|}{-56.13\%}         \\
        \multicolumn{1}{|c|}{} & 250 & 50 & 8  & 27.68 & \multicolumn{1}{c|}{6920} & 38.79 & \multicolumn{1}{c|}{-83.57\%} & 29.90 & \multicolumn{1}{c|}{-55.74\%}         \\
        \multicolumn{1}{|c|}{} & 230 & 50 & 11 & 29.97 & \multicolumn{1}{c|}{6893} & 36.91 & \multicolumn{1}{c|}{-84.43\%} & 28.03 & \multicolumn{1}{c|}{-58.67\%}         \\
        \multicolumn{1}{|c|}{} & 190 & 50 & 14 & 36.31 & \multicolumn{1}{c|}{6899} & 32.65 & \multicolumn{1}{c|}{-86.22\%} & 23.77 & \multicolumn{1}{c|}{-64.92\%}         \\
        \multicolumn{1}{|c|}{} & 210 & 65 & 8 & 34.18  & \multicolumn{1}{c|}{7178} & 34.48 & \multicolumn{1}{c|}{-84.86\%} & 25.60 & \multicolumn{1}{c|}{-60.70\%}         \\
        \multicolumn{1}{|c|}{} & 190 & 65 & 11 & 37.08 & \multicolumn{1}{c|}{7045} & 32.91 & \multicolumn{1}{c|}{-85.81\%} & 24.03 & \multicolumn{1}{c|}{-63.79\%}         \\
        \multicolumn{1}{|c|}{} & 170 & 65 & 14 & 52.37 & \multicolumn{1}{c|}{8903} & 27.15 & \multicolumn{1}{c|}{-85.21\%} & 18.27 & \multicolumn{1}{c|}{-65.21\%}   \\ \hdashline
        \multicolumn{1}{|c|}{\multirow{9}{*}{Night   Charging (NC)}} & 260 & 35 & 8 & 28.53 & \multicolumn{1}{c|}{7418} & 37.34 & \multicolumn{1}{c|}{-83.05\%} & 28.46 & \multicolumn{1}{c|}{-54.84\%}         \\
        \multicolumn{1}{|c|}{} & 260 & 35 & 11 & 27.63 & \multicolumn{1}{c|}{7184} & 38.06 & \multicolumn{1}{c|}{-83.27\%} & 29.17 & \multicolumn{1}{c|}{-55.18\%}         \\
        \multicolumn{1}{|c|}{} & 240 & 35 & 14 & 33.7  & \multicolumn{1}{c|}{8088} & 33.63 & \multicolumn{1}{c|}{-83.36\%} & 24.74 & \multicolumn{1}{c|}{-57.20\%}         \\
        \multicolumn{1}{|c|}{} & 240 & 50 & 8  & 29.49 & \multicolumn{1}{c|}{7078} & 37.13 & \multicolumn{1}{c|}{-83.92\%} & 28.25 & \multicolumn{1}{c|}{-57.23\%}         \\
        \multicolumn{1}{|c|}{} & 210 & 50 & 11 & 37.21 & \multicolumn{1}{c|}{7814} & 32.23 & \multicolumn{1}{c|}{-84.59\%} & 23.34 & \multicolumn{1}{c|}{-60.99\%}         \\
        \multicolumn{1}{|c|}{} & 210 & 50 & 14 & 38.27 & \multicolumn{1}{c|}{8037} & 31.68 & \multicolumn{1}{c|}{-84.42\%} & 22.80 & \multicolumn{1}{c|}{-60.81\%}         \\
        \multicolumn{1}{|c|}{} & 200 & 65 & 8  & 35.28 & \multicolumn{1}{c|}{7056} & 33.82 & \multicolumn{1}{c|}{-85.40\%} & 24.94 & \multicolumn{1}{c|}{-62.36\%}         \\
        \multicolumn{1}{|c|}{} & 190 & 65 & 11 & 40.67 & \multicolumn{1}{c|}{7727} & 31.19 & \multicolumn{1}{c|}{-85.25\%} & 22.30 & \multicolumn{1}{c|}{-63.14\%}         \\
        \multicolumn{1}{|c|}{} & 180 & 65 & 14 & 41.92 & \multicolumn{1}{c|}{7546} & 30.60 & \multicolumn{1}{c|}{-85.87\%} & 21.72 & \multicolumn{1}{c|}{-64.94\%}  \\  \hdashline
        \multicolumn{1}{|c|}{\multirow{9}{*}{Strategic Dispatching (SD)}} & 300 & 35 & 8 & 24.76 & \multicolumn{1}{c|}{7428} & 41.08 & \multicolumn{1}{c|}{-81.33\%} & 32.20 & \multicolumn{1}{c|}{-48.84\%}         \\
        \multicolumn{1}{|c|}{} & 280 & 35 & 11 & 26.26 & \multicolumn{1}{c|}{7353} & 39.35 & \multicolumn{1}{c|}{-82.30\%} & 30.46 & \multicolumn{1}{c|}{-52.09\%}         \\
        \multicolumn{1}{|c|}{} & 270 & 35 & 14 & 28.46 & \multicolumn{1}{c|}{7684} & 37.34 & \multicolumn{1}{c|}{-82.44\%} & 28.46 & \multicolumn{1}{c|}{-53.22\%}         \\
        \multicolumn{1}{|c|}{} & 240 & 50 & 8  & 29.31 & \multicolumn{1}{c|}{7034} & 37.36 & \multicolumn{1}{c|}{-83.92\%} & 28.47 & \multicolumn{1}{c|}{-57.16\%}         \\
        \multicolumn{1}{|c|}{} & 230 & 50 & 11 & 33.70 & \multicolumn{1}{c|}{7751} & 34.23 & \multicolumn{1}{c|}{-83.76\%} & 25.34 & \multicolumn{1}{c|}{-57.99\%}         \\
        \multicolumn{1}{|c|}{} & 230 & 50 & 14 & 30.28 & \multicolumn{1}{c|}{6964} & 36.49 & \multicolumn{1}{c|}{-84.45\%} & 27.60 & \multicolumn{1}{c|}{-58.89\%}         \\
        \multicolumn{1}{|c|}{} & 180 & 65 & 8  & 40.90 & \multicolumn{1}{c|}{7362} & 31.07 & \multicolumn{1}{c|}{-86.00\%} & 22.18 & \multicolumn{1}{c|}{-65.07\%}         \\
        \multicolumn{1}{|c|}{} & 180 & 65 & 11 & 41.70 & \multicolumn{1}{c|}{7506} & 30.72 & \multicolumn{1}{c|}{-85.89\%} & 21.83 & \multicolumn{1}{c|}{-64.95\%}         \\
        \multicolumn{1}{|c|}{} & 180 & 65 & 14 & 39.74 & \multicolumn{1}{c|}{7153} & 31.50 & \multicolumn{1}{c|}{-88.21\%} & 22.67 & \multicolumn{1}{c|}{-65.31\%} \\ \hdashline
        \multicolumn{1}{|c|}{\multirow{9}{*}{Fast Charging (FD)}} & 150 & 35 & 8 & 46.3 & \multicolumn{1}{c|}{6945} & 29.11 & \multicolumn{1}{c|}{-87.63\%} & 20.22 & \multicolumn{1}{c|}{-69.96\%}         \\
        \multicolumn{1}{|c|}{} & 110 & 35 & 11 & 66.98 & \multicolumn{1}{c|}{7368} & 24.25 & \multicolumn{1}{c|}{-89.07\%} & 15.36 & \multicolumn{1}{c|}{-75.79\%}         \\
        \multicolumn{1}{|c|}{} & 90  & 35 & 14 & 86.11 & \multicolumn{1}{c|}{7750} & 21.81 & \multicolumn{1}{c|}{-89.66\%} & 12.92 & \multicolumn{1}{c|}{-78.58\%}         \\
        \multicolumn{1}{|c|}{} & 140 & 50 & 8  & 52.7  & \multicolumn{1}{c|}{7378} & 27.98 & \multicolumn{1}{c|}{-87.37\%} & 19.10 & \multicolumn{1}{c|}{-69.86\%}         \\
        \multicolumn{1}{|c|}{} & 110 & 50 & 11 & 69.2  & \multicolumn{1}{c|}{7612} & 24.46 & \multicolumn{1}{c|}{-88.61\%} & 15.56 & \multicolumn{1}{c|}{-74.67\%}         \\
        \multicolumn{1}{|c|}{} & 90  & 50 & 14 & 83.95 & \multicolumn{1}{c|}{7556} & 22.51 & \multicolumn{1}{c|}{-89.59\%} & 13.63 & \multicolumn{1}{c|}{-77.97\%}         \\
        \multicolumn{1}{|c|}{} & 150 & 65 & 8  & 46.13 & \multicolumn{1}{c|}{6920} & 30.96 & \multicolumn{1}{c|}{-86.89\%} & 22.07 & \multicolumn{1}{c|}{-67.34\%}         \\
        \multicolumn{1}{|c|}{} & 110 & 65 & 11 & 67.38 & \multicolumn{1}{c|}{7412} & 25.36 & \multicolumn{1}{c|}{-88.50\%} & 16.47 & \multicolumn{1}{c|}{-73.89\%}         \\
        \multicolumn{1}{|c|}{} & 90  & 65 & 14 & 82.52 & \multicolumn{1}{c|}{7427} & 23.15 & \multicolumn{1}{c|}{-89.48\%} & 14.27 & \multicolumn{1}{c|}{-77.33\%}         \\ \hline
        \end{tabular}
        \caption{Results showing the fleet size needed to cover the demand with the desired service requirements, total system kilometers traveled, grams of $\mathrm{CO_2}$ per km traveled and $\mathrm{CO_2}$ reductions compared to the baseline scenario.}%
        \label{tab:Environmental impacts}
    }
\end{table}

In contrasting the existing car-based baseline scenario with the shared lightweight autonomous vehicle-based scenario, a consistently positive environmental impact is observed across all assessed vehicle configurations and charging strategies. Notably, as detailed in Table \ref{tab:Environmental impacts}, under the current US electricity composition, transitioning food deliveries from combustion car-based to shared lightweight autonomous vehicle-based approach showcases potential for reducing $\mathrm{CO_2}$ equivalent emissions between 81.33 \% and 89.66\%. Furthermore, even when contemplating a prospective car-based scenario utilizing BEVs in conjunction with a zero-carbon electricity grid, the reductions remain significant, ranging from 48.34\% to 78.58\%.

A recurring pattern emerges regarding the implications of the distinct \textit{autonomous driving speeds} explored within this study: elevating the speed typically corresponds to a greater reduction in $\mathrm{CO_2}$ emissions. Specifically, the reduction is particularly pronounced when transitioning from a slow configuration to a medium-speed variant, compared to the shift from medium to high speed. This trend is inherently tied to the system's minimal performance-driven fleet size requisites. With increased vehicle speed, the time required to serve each trip is lower, subsequently decreasing the necessary fleet size to meet demand. Despite the potential increase in total kilometers traveled due to fewer vehicles, this drawback is counterbalanced by the reduction in environmental impact resulting from a decreased fleet size.

Concerning \textit{battery ranges}, in scenarios in which conventional charging is employed (CC, NC, and SD), a bigger battery range is correlated with a smaller fleet size, thereby reducing the environmental impact. Since the fleet size reduction is higher in changing from small to medium batteries than in changing medium to large batteries (see Section \ref{subsubsec:Vehicle configurations}), the environmental impact reductions are also more considerable from small to medium batteries. However, a divergent trend emerges in fast charging (FC) scenarios: a larger battery range yields a smaller reduction in $\mathrm{CO_2}$ emissions. As the charging events in the FC scenario do not influence fleet sizing, increasing the battery range results in larger and more environmentally intensive batteries, with no subsequent reduction in fleet sizes. For this reason, increasing the battery range in the FC scenario has a negative impact.

Finally, considering the impact of the different \textit{charging operational strategies}, it is concluded that the reductions in the CC, NC, and SD scenarios are similar, while the FC scenario exhibits the most substantial reductions. Consequently, despite the necessity for twice as many batteries per vehicle in the FC scenario, the resultant reduction in fleet size contributes to reducing the environmental impact.

In conclusion, while shared lightweight autonomous systems require more vehicles than current car-based systems, this study underscores their potential to effectively and significantly mitigate environmental impacts. Moreover, results highlight a high dependency of the environmental impacts on the configuration metrics. This indicates that fleet design and operation-related decisions can have a determinant effect. Therefore, decision-making processes and regulations can play a significant role in defining the environmental outcomes of lightweight autonomous vehicles for food deliveries.

\section{Interactive simulation tool}
\label{sec:Tangible interface}

Given the significance of the findings from the simulation study detailed in this article, an interactive version of the simulation model was developed to provide stakeholders with a firsthand exploration of the model's outcomes. The model is dynamically linked to the ABM and environmental impact study outlined in Section \ref{sec:Modeling}, enabling users to interact with the design variables discussed in Section \ref{sec:Experimental setup} and receive real-time feedback.

\begin{figure}[!h]
    \centering
    \caption{Information on the scenario (1 – red), environmental impacts (2 – green), service level (3 – blue), and fleet performance (4 – yellow) display that the users analyze to generate insights.}
    \includegraphics[width=0.5\linewidth]{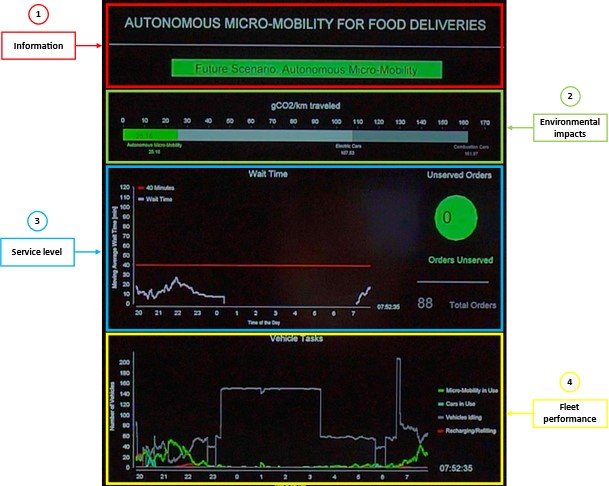}
    \label{fig:Indicators display}
\end{figure}

\begin{figure}[!h]
    \centering
    \caption{Interactive board (1 – red), legend (2 – green), road network (3 – blue), and agents (4 – yellow) displays that the user needs to analyze to generate insights.}
    \includegraphics[width=0.5\linewidth]{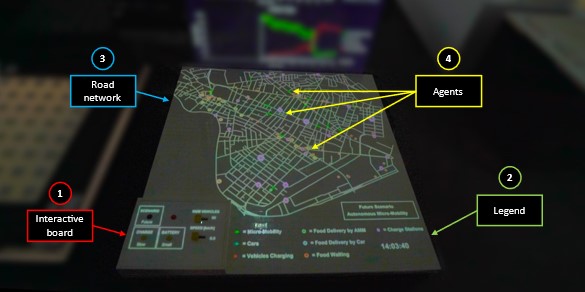}
    \label{fig:Map display}
\end{figure}

Inspired by projects like the CityScope, we have developed a tangible tool that allows for collaborative manipulation by different users \citep{noyman2022cityscope,alonso2018cityscope}. The configuration comprises two main components: 1) a television monitor showcasing various performance indicators (Figure \ref{fig:Indicators display}), and 2) a dynamic map display projected onto a horizontal surface (Figure \ref{fig:Map display}).

\begin{figure}[!h]
    \centering
    \caption{Schema depicting the different scenarios and parameters users can explore in the interactive simulation tool.}
    \includegraphics[width=0.4\linewidth]{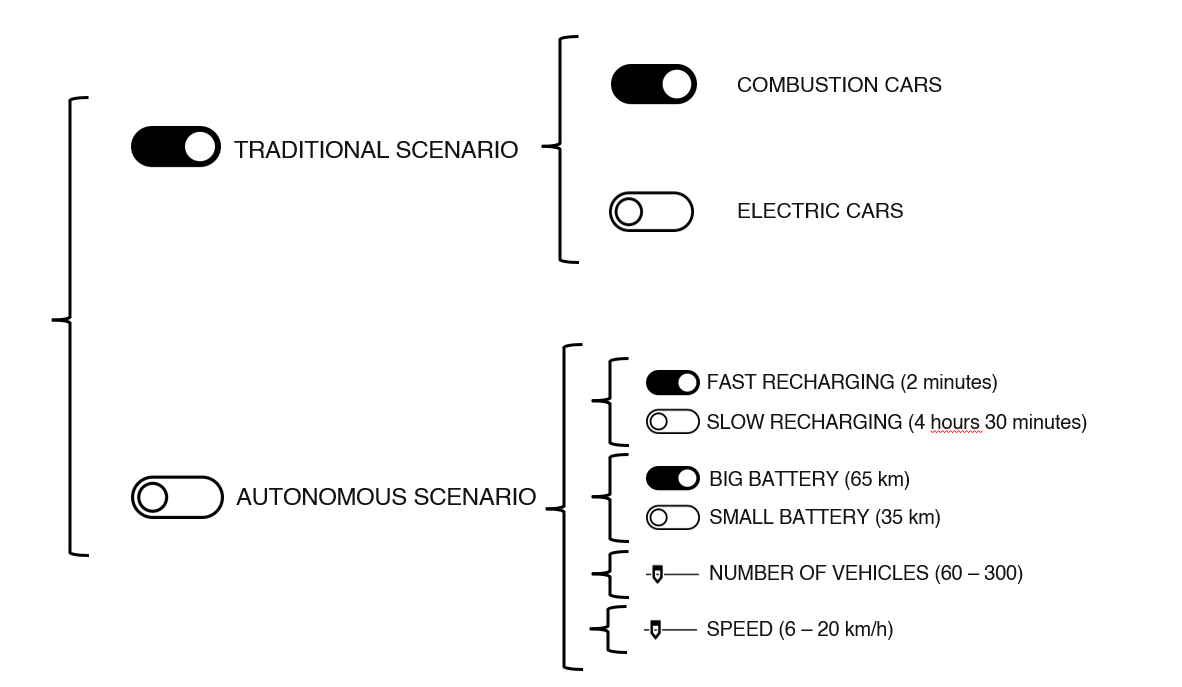}
    \label{fig:Interactions}
\end{figure}

Users engage with the model through a custom-made interactive board (Figure \ref{fig:Map display} - 1). This board allows users to select the simulation scenarios and define their lightweight autonomous vehicle fleet system, customizing vehicle configurations and charging strategies, as depicted in Figure \ref{fig:Interactions}.

The indicators' display, illustrated in Figure \ref{fig:Indicators display}, offers insights into environmental impacts, service levels, and fleet performance. It begins with an informative category showcasing the project's title and visualized scenarios. Environmental impacts are presented through a dynamic bar chart depicting $\mathrm{CO_2}$ emissions per vehicle kilometer, updating in real-time according to user interactions. Service level indicators are comprised of an average wait time chart for food delivery orders and a counter for unfulfilled orders. Additionally, real-time vehicle activities are showcased, representing the tasks (idle, pick up/delivery, charging) that different vehicles are doing and how they evolve over time.

Lastly, the dynamic map display in Figure \ref{fig:Map display} provides a detailed visual overview of the study area's map, agents' movements on roads, and relevant simulation activities. This display incorporates: a city road network map, moving agents with real-time activities represented using distinct colors and shapes, and an accompanying legend.

As mobility systems grow in complexity, so does decision-making. This interactive and adaptable foster consensus in complex, multi-stakeholder scenarios. By facilitating meaningful discussions, it could help stakeholders grasp trade-offs and perspectives, ultimately informing better decisions in mobility system design, development, and deployment.  We have done preliminary evaluations of the effectiveness of this tool with diverse stakeholders. However, we anticipate conducting more comprehensive testing in the future to evaluate its usefulness in real decision-making processes.

\section{Conclusions}
\label{sec:Conclusions}

This study focuses on the mobility innovations that have been catalyzed by the surge in online food deliveries in recent years. As researchers and delivery companies explore lightweight autonomous vehicles to serve food deliveries, this research focuses on the fleet-level performance and environmental implications of these new vehicles. We assess the impact of diverse autonomous vehicle configurations and charging strategies in fleet-level performance through an agent-based model, and we evaluate the corresponding environmental implications through a life cycle assessment. 

The findings of our analysis reveal that driving speed and battery range influence fleet size, with faster speeds and extended ranges leading to reduced fleet requirements. Charging strategies exhibit diverse impacts on fleet size, with fast charging proving to be the most efficient in reducing fleet sizes and environmental impacts. Overall, the potential for substantial environmental mitigation is evident despite the larger fleet sizes required for autonomous systems compared to current car-based services. These conclusions are a significant step in evaluating the viability of lightweight autonomous vehicles as a transformative alternative to conventional food delivery practices. 

Lastly, the interactive decision-making tool developed offers stakeholders a user-friendly platform to extract valuable insights and facilitate informed discussions, supporting decision-making within this evolving landscape.



\section{Declaration of interest}

Authors have no conflict of interest to declare.

 \bibliographystyle{elsarticle-harv} 
 \bibliography{cas-refs}

\begin{appendices}
\section{}
\label{app:Tables and Figures}
\begin{figure}[!h]
    \centering
    \caption{Complete diagram of the synthetic database generation process for obtaining the fine-grained food delivery demand dataset.}
    \includegraphics[width=0.9\linewidth]{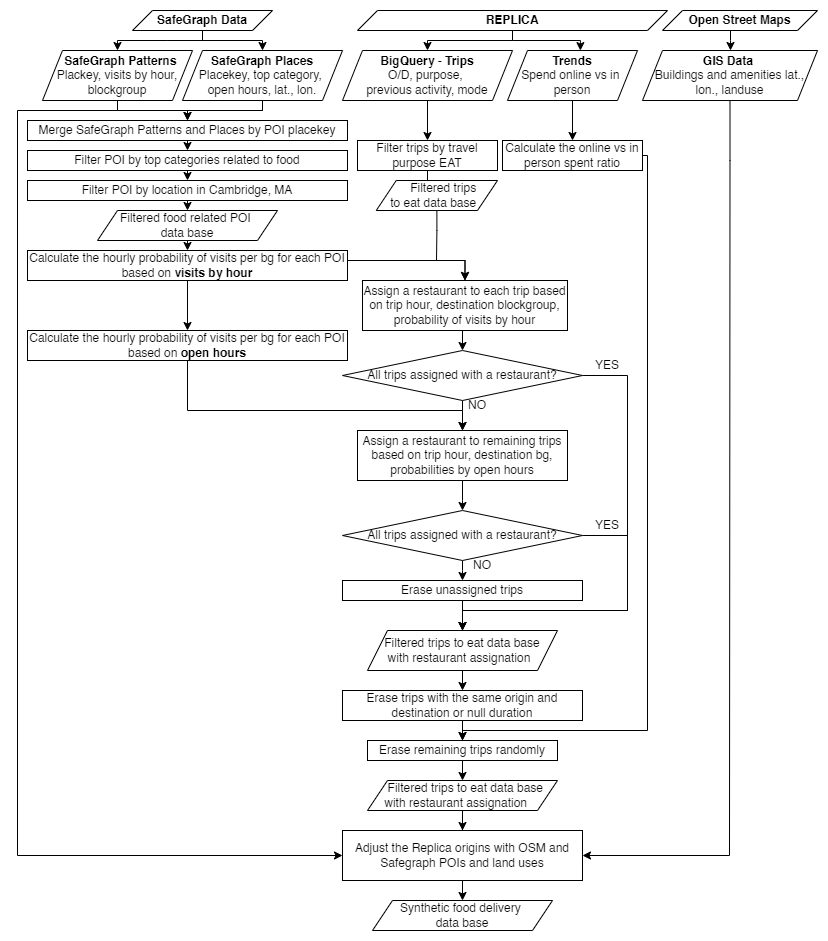}
    \label{fig:Synthetic Database Generation}
\end{figure}

\begin{table}[!h]

\resizebox{0.8\textwidth}{!}{%
    \centering
    \caption{Land use tag association to each of the Replica trips based on their origin building land use information.}
    \label{tab:Land use association Replica} 

\begin{tabular}{|l|p{12cm}|}
\hline
\textbf{LAND USE} & \multicolumn{1}{c|}{\textbf{ORIGIN BUILDING LAND USE}} \\ \hline
Residential & \begin{tabular}[c]{@{}l@{}}Residential\\  Mixed-use – Residential\\  Mixed-use – Retail – Previous activity = Home or Work from home\end{tabular} \\ \hline
Industrial & \begin{tabular}[c]{@{}l@{}}Industrial\\  Mixed-use – Industrial\end{tabular} \\ \hline
Office & \begin{tabular}[c]{@{}l@{}}Mixed-use – Office\\  Commercial – Office\end{tabular} \\ \hline
Shop & \begin{tabular}[c]{@{}l@{}}Mixed-use – Retail – Previous activity = Shop\\  Commercial – Retail – Previous activity = Shop\end{tabular} \\ \hline
Hotel & Commercial – Retail – Previous activity = Lodging \\ \hline
Restaurants & Mixed-use – Retail – Previous activity = Eat, Social or Other Activity Commercial – Retail – Previous activity = Eat, Social or Other Activity \\ \hline
Work & \begin{tabular}[c]{@{}l@{}}Mixed-use – Retail – Previous activity = Work or Maintenance\\  Commercial – Retail – Previous activity = Work or Maintenance\end{tabular} \\ \hline
Non-Retail Attractions & \begin{tabular}[c]{@{}l@{}}Mixed-use – Non-retail attraction\\  Commercial – Non-retail attraction\end{tabular} \\ \hline
Park & \begin{tabular}[c]{@{}l@{}}Open space\\  Mixed-use – Open Space\\  Mixed-use – Retail – Previous activity = Recreation\\  Commercial – Retail – Previous activity = Recreation\end{tabular} \\ \hline
Transportation Utilities & \begin{tabular}[c]{@{}l@{}}Transportation utilities\\  Mixed-use – Transportation utilities\end{tabular} \\ \hline
Civic Institutional & \begin{tabular}[c]{@{}l@{}}Civic institutional – Civic institutional\\  Mixed-use – Civic institutional\end{tabular} \\ \hline
Education & \begin{tabular}[c]{@{}l@{}}Civic-institutional – Education \\  Mixed-use – Education\\  Mixed-use – Retail – Previous Activity = School\\  Commercial – Retail – Previous Activity = School\end{tabular} \\ \hline
Healthcare & \begin{tabular}[c]{@{}l@{}}Civic-institutional – Healthcare \\  Mixed-use – Healthcare\end{tabular} \\ \hline
\end{tabular}
}
\end{table}

\begin{table} 
\resizebox{0.8\textwidth}{!}{%
\caption{Land use tag association to each of the SafeGraph and Open Street Map points of interest (Part 1/4). } 
\label{tab:Land use association SafeGraph} 
\begin{tabular}{|c|l|}
\hline
\textbf{LAND USE} & \multicolumn{1}{c|}{\textbf{ORIGIN BUILDING LAND USE}} \\ \hline
Residential & \begin{tabular}[c]{@{}l@{}}Open Street Map: \\  - Residential.\\  - Mixed-use residential.\\  - Assisted living/boarding house.\\  - Education residential.\end{tabular} \\ \hline
Industrial & \begin{tabular}[c]{@{}l@{}}SAFEGRAPH:\\  - Gambling Industries.\\  - Coating, engraving, heat treating, and allied activities.\\  - Machinery, equipment, and supplies merchant wholesalers.\\  - Motion picture and video industries.\\  - Converted paper product manufacturing.\\  - Glass and glass product manufacturing.\\  - Electric power generation, transmission, and distribution.\\  - Beverage manufacturing.\\  - Sound recording industries.\\  - Bakeries and tortilla manufacturing.\\  - Other amusement and recreation industries.\\  - Other miscellaneous manufacturing.\end{tabular} \\ \hline
Office & \begin{tabular}[c]{@{}l@{}}SAFEGRAPH:\\  - Offices of real estate agents and brokers.\\  - Management of companies and enterprises.\\  - Agencies, brokerages, and other insurance-related activities.\\  - Electronic and precision equipment repair and maintenance.\\  - Architectural, engineering, and related services.\\  - Personal and household goods repair and maintenance.\\  - Other professional, scientific, and technical assistance.\\  - Building equipment contractors.\\  - Automobile dealers.\\  - Activities related to real estate.\\  - Other financial investment activities.\\  - Travel arrangement and reservation services.\\  - Radio and television broadcasting.\\  - Automotive equipment rental and leasing.\\  - Management, scientific, and technical consulting services.\\  - Building materials and supplies dealers.\\  - Consumer goods rental.\\  - Building finishing contractors.\\  - Couriers and express delivery services.\\  - Cable and other subscription programming.\\  - Advertising, public relations, and related services.\\  - Administration of human resource programs.\\  - Other specialty trade contractors.\end{tabular} \\ \hline
\end{tabular}
}
\end{table}

\begin{table}
\resizebox{0.9\textwidth}{!}{%
\caption{Land use tag association to each of the SafeGraph and Open Street Map points of interest (Part 2/4).}
\begin{tabular}{|c|l|}
\hline
Shop & \begin{tabular}[c]{@{}l@{}}SAFEGRAPH:\\  - Jewelry, luggage, and leather goods stores.\\  - Clothing stores.\\  - Office supplies, stationery, and gift stores.\\  - Furniture stores.\\  - Beer, wine, and liquor stores.\\  - Grocery stores.\\  - Specialty food store.\\  - Shoe stores.\\  - Florists.\\  - Health and personal care stores.\\  - Printing and related support activities.\\  - Home furnishing stores.\\  - Electronics and appliance stores.\\  - Department stores.\\  - Used merchandise stores.\\  - Drugs and druggists’ sundries merchant wholesalers.\\  - General merchandise stores, including warehouse clubs and supercenters.\\  - Lawn and garden equipment and supplies stores.\\  - Other motor vehicle dealers.\\  - Book stores and new dealers.\\  - Other miscellaneous store retailers.\end{tabular} \\ \hline
Hotel & \begin{tabular}[c]{@{}l@{}}SAFEGRAPH:\\  - Traveler accommodation.\end{tabular} \\ \hline
Restaurants & \begin{tabular}[c]{@{}l@{}}SAFEGRAPH:\\  - Restaurants and other eating places.\\  - Drinking places (alcoholic beverages).\\  - Special food services.\end{tabular} \\ \hline
Work & SAFEGRAPH: All the shops, hotels, and restaurants \\ \hline
Non-Retail Attractions & \begin{tabular}[c]{@{}l@{}}SAFEGRAPH:\\  - Sporting goods, hobbies, and musical instrument stores.\\  - Museums, historical sites, and similar instructions.\\  - Amusement parks and arcades.\\  - Performing arts companies.\\  - Promoters of performing arts, sports, and similar events.\\  - Religious organizations.\end{tabular} \\ \hline
Park & \begin{tabular}[c]{@{}l@{}}Open Street Map:\\  - Public open space.\\  - Private own open space.\end{tabular} \\ \hline
\end{tabular}
}
\end{table}

\begin{table}
\resizebox{0.8\textwidth}{!}{%
\caption{Land use tag association to each of the SafeGraph and Open Street Map points of interest (Part 3/4). }
\begin{tabular}{|c|l|}
\hline
Transportation Utilities & \begin{tabular}[c]{@{}l@{}}Open Street Map:\\  - Bicycle parking.\\  - Bicycle repair station.\\  - Bicycle rental.\\  - Boat rental.\\  - Boat sharing.\\  - Bus station.\\  - Car rental.\\  - Car sharing.\\  - Car wash.\\  - Vehicle inspection.\\  - Charging station.\\  - Ferry terminal.\\  - Fuel.\\  - Grit bin.\\  - Motorcycle parking.\\  - Parking.\\  - Parking entrance.\\  - Parking space.\\  - Taxi.\end{tabular} \\ \hline
Civic Institutional & \begin{tabular}[c]{@{}l@{}}Open Street Map:\\  - Courthouse.\\  - Embassy.\\  - Fire station.\\  - Police.\\  - Post box.\\  - Post depot.\\  - Post office.\\  - Prison.\\  - Ranger station.\\  - Townhall.\end{tabular} \\ \hline
Education & \begin{tabular}[c]{@{}l@{}}SAFEGRAPH:\\  - Colleges, universities, and professional schools.\\  - Technical trade and trade schools.\\  - Administration of economic programs.\\  - Elementary and secondary schools.\\  - Child day care services.\\  - Other schools and instruction.\end{tabular} \\ \hline
\end{tabular}
}
\end{table}

\begin{table}
\resizebox{0.8\textwidth}{!}{%
\caption{Land use tag association to each of the SafeGraph and Open Street Map points of interest (Part 4/4).}
\begin{tabular}{|c|l|}
\hline
Healthcare & \begin{tabular}[c]{@{}l@{}}SAFEGRAPH:\\ - Offices of physicians.\\ - Offices of dentists.\\ - Offices of other health practitioners.\\ - Outpatient care centers.\\ - Nursing care facilities (skilled nursing facilities).\\ - Nursing and residential care facilities.\\ - Medical and diagnostic laboratories.\\ - General medical and surgical hospitals.\\ - Specialty (except psychiatric and substance abuse) hospitals.\\ - Insurance carriers.\\ - Personal care services.\\ - Individual and family services.\\ - Death care services.\\ - Other personal services.\\ - Home health care services.\end{tabular} \\ \hline
\end{tabular}
}
\end{table}
\begin{figure}[!h]
        \captionsetup{skip=0pt}
        \caption{Number of SLAV carrying out different activities throughout the day in the conventional charging (CC) sub-scenario.}
        \includegraphics[width=0.8\linewidth]{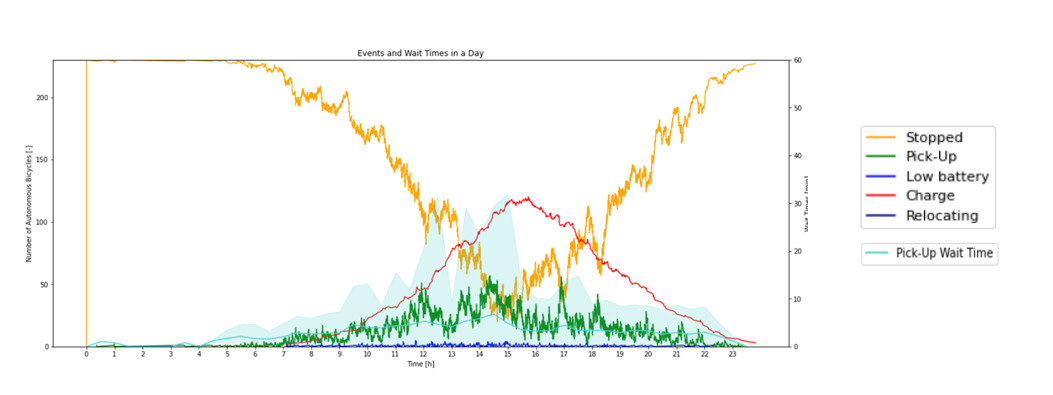}
        \label{fig:Vehicles state CC}
\end{figure}
\begin{figure}[!h]
        \captionsetup{skip=0pt}
        \caption{Number of SLAV carrying out different activities throughout the day in the night charging (NC) sub-scenario.}
        \includegraphics[width=0.8\linewidth]{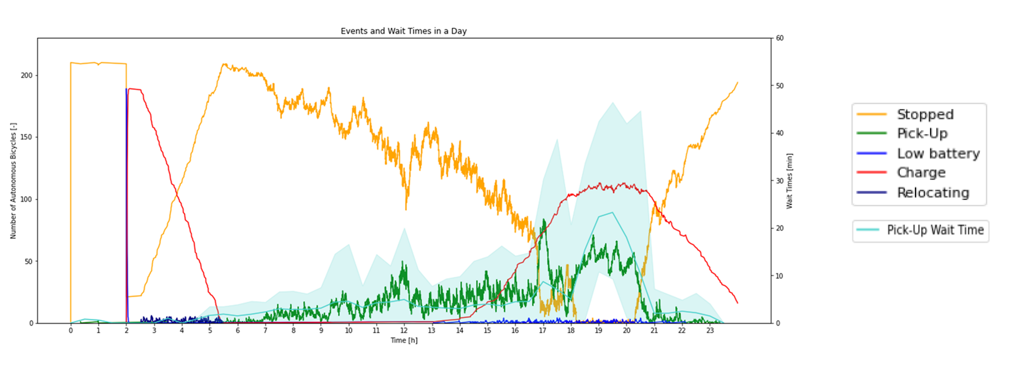}
        \label{fig:Vehicles state NC}
\end{figure}
\begin{figure}[!h]
        \captionsetup{skip=0pt}
        \caption{Number of SLAV carrying out different activities throughout the day in the strategic dispatching (SD) sub-scenario.}
        \includegraphics[width=0.8\linewidth]{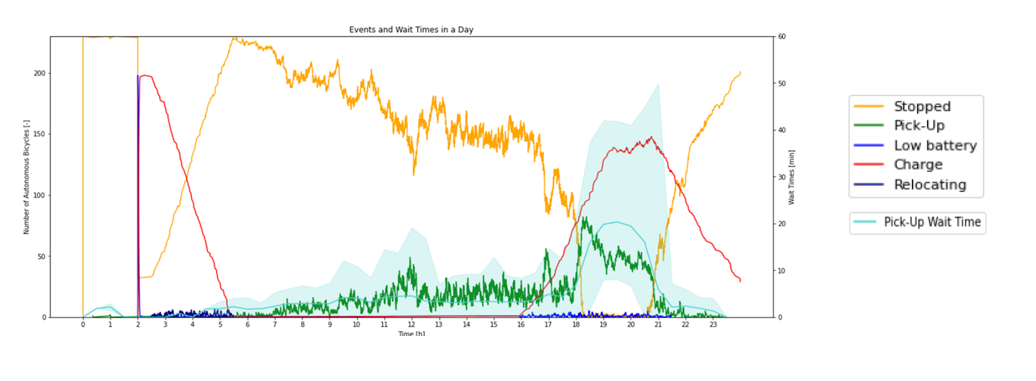}
        \label{fig:Vehicles state SD}
\end{figure}
\begin{figure}[!h]
        \captionsetup{skip=0pt}
        \caption{Number of SLAV carrying out different activities throughout the day in the fast charging (FC) sub-scenario.}
        \includegraphics[width=0.8\linewidth]{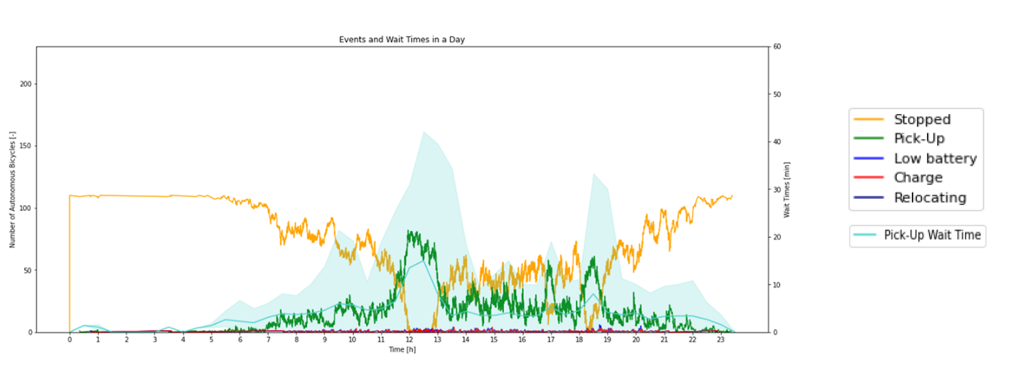}
        \label{fig:Vehicles state FC}
\end{figure}

\end{appendices}
\end{document}